\documentclass[times,11pt,a4paper]{article}

\usepackage{epsfig}
\usepackage{graphicx}
\usepackage{amsmath}
\usepackage{amssymb}
\usepackage{amsthm}
\usepackage{float}
\usepackage{url}

\begin{document}

\setlength{\leftmargini}{0\leftmargini}
\newtheorem{definition}{Definition}
\newtheorem{lemma}{Lemma}
\newtheorem{proposition}{Proposition}
\newtheorem{corollary}{Corollary}
\newtheorem{theorem}{Theorem}
\newtheorem{conjecture}{Conjecture}
\newtheorem{remark}{Remark}
\newtheorem{example}{Example}
\newtheorem{discussion}{Discussion}

\newcommand{\dsum}{\displaystyle\sum}
\newcommand{\dfr}{\displaystyle\frac}
\newcommand{\dint}{\displaystyle\int}
\newcommand{\dprod}{\displaystyle\prod}
\newcommand{\naturals}{\ensuremath{\mathbb{N}}}
\newcommand{\reals}{\ensuremath{\mathbb{R}}}
\newcommand{\expectation}{\ensuremath{\mathbb{E}}}
\newcommand{\vect}{\mathbf}

\newcommand\T{\rule{0pt}{2.6ex}}
\newcommand\B{\rule[-1.2ex]{0pt}{0pt}}

\title{\bf{Bounds on the Number of Iterations
for Turbo-Like Ensembles over the Binary Erasure Channel}}

\date{}
\maketitle

\vspace{-1.5cm}
\begin{center}
\begin{tabular}{cc c cc}
Igal Sason \quad \quad \quad Gil Wiechman\\
Technion -- Israel Institute of technology\\
Haifa 32000, Israel\\
{\tt \{sason@ee, igillw@tx\}.technion.ac.il}
\end{tabular}
\end{center}

\hspace{6.1cm} \today

\begin{abstract}
This paper provides simple lower bounds on the number of
iterations which is required for successful message-passing
decoding of some important families of graph-based code ensembles
(including low-density parity-check codes and variations of
repeat-accumulate codes). The transmission of the code ensembles
is assumed to take place over a binary erasure channel, and the
bounds refer to the asymptotic case where we let the block length
tend to infinity. The simplicity of the bounds derived in this
paper stems from the fact that they are easily evaluated and are
expressed in terms of some basic parameters of the ensemble which
include the fraction of degree-2 variable nodes, the target bit
erasure probability and the gap between the channel capacity and
the design rate of the ensemble. This paper demonstrates that the
number of iterations which is required for successful
message-passing decoding scales at least like the inverse of the
gap (in rate) to capacity, provided that the fraction of degree-2
variable nodes of these turbo-like ensembles does not vanish
(hence, the number of iterations becomes unbounded as the gap to
capacity vanishes).
\end{abstract}


\section{Introduction}
\label{Section: Introduction} During the last decade, there have
been many developments in the construction and analysis of
low-complexity error-correcting codes which closely approach the
Shannon capacity limit of many standard communication channels
with feasible complexity. These codes are understood to be codes
defined on graphs, together with the associated iterative decoding
algorithms. Graphs serve not only to describe the codes
themselves, but more importantly, they structure the operation of
their efficient sub-optimal iterative decoding algorithms.

Proper design of codes defined on graphs enables to asymptotically
achieve the capacity of the binary erasure channel (BEC) under
iterative message-passing decoding. Capacity-achieving sequences
of ensembles of low-density parity-check (LDPC) codes were
originally introduced by Shokrollahi \cite{Shokrollahi-IMA2000}
and by Luby et al. \cite{LubyMSS_IT01}, and a systematic study of
capacity-achieving sequences of LDPC ensembles was presented by
Oswald and Shokrollahi \cite{Oswald-it02} for the BEC. Analytical
bounds on the maximal achievable rates of LDPC ensembles were
derived by Barak et al. \cite{Barak_IT04} for the asymptotic case
where the block length tends to infinity; this analysis provides a
lower bound on the gap between the channel capacity and the
achievable rates of LDPC ensembles under iterative decoding. The
decoding complexity of LDPC codes under iterative message-passing
decoding scales linearly with the block length, though their
encoding complexity is in general super-linear with the block
length; this motivated the introduction of repeat-accumulate codes
and their more recent variants (see, e.g., \cite{Abbasfar_Comm07},
\cite{McEliece_IRA_codes} and \cite{Pfister_Sason_IT07}) whose
encoding and decoding complexities under iterative message-passing
decoding are both inherently linear with the block length. Due to
the simplicity of the density evolution analysis for the BEC,
suitable constructions of capacity-achieving ensembles of variants
of repeat-accumulate codes were devised in
\cite{McEliece_IRA_codes}, \cite{PfisterSU_IT05},
\cite{Pfister_Sason_IT07} and \cite{Sason-it04}. All these works
rely on the density evolution analysis of codes defined on graphs
for the BEC, and provide an asymptotic analysis which refers to
the case where we let the block length of these code ensembles
tend to infinity. Another innovative coding technique, introduced
by Shokrollahi \cite{Shokroallhi-IT06}, enables to achieve the
capacity of the BEC with encoding and decoding complexities which
scale linearly with the block length, and it has the additional
pleasing property of achieving the capacity without the knowledge
of the erasure probability of the channel.

The performance analysis of finite-length LDPC code ensembles
whose transmission takes place over the BEC was introduced by Di
et al. \cite{Di_IT02}. This analysis considers sub-optimal
iterative message-passing decoding as well as optimal
maximum-likelihood decoding. In \cite{Abdel_ETT07}, an efficient
approach to the design of LDPC codes of finite length was
introduced by Amraoui et al.; this approach is specialized for the
BEC, and it enables to design such code ensembles which perform
well under iterative decoding with a practical constraint on the
block length. In \cite{Richardson_Urbanke_unpublished}, Richardson
and Urbanke initiated the analysis of the distribution of the
number of iterations needed for the decoding of LDPC ensembles of
finite block length which are communicated over the BEC.

For general channels, the number of iterations is an important
factor in assessing the decoding complexity of graph-based codes
under iterative message-passing decoding. The second factor
determining the decoding complexity of such codes is the
complexity of the Tanner graph which is used to represent the
code; this latter quantity, defined as the number of edges in the
graph per information bit, serves as a measure for the decoding
complexity per iteration.

The extrinsic information transfer (EXIT) charts, pioneered by
Stephan ten Brink \cite{EXIT_paper_Comm01, EXIT_paper_Annals01},
form a powerful tool for an efficient design of codes defined on
graphs by tracing the convergence behavior of their iterative
decoders. EXIT charts provide a good approximative engineering
tool for tracing the convergence behavior of soft-input
soft-output iterative decoders; they suggest a simplified
visualization of the convergence of these decoding algorithms,
based on a single parameter which represents the exchange of
extrinsic information between the constituent decoders. For the
BEC, the EXIT charts coincide with the density evolution analysis
(see \cite{Richardson_Urbanke_book}) which is simplified in this
case to a one-dimensional analysis.

A numerical approach for the joint optimization of the design rate
and decoding complexity of LDPC ensembles was provided in
\cite{ArdakaniSFK_Allerton05}; it is assumed there that the
transmission of these code ensembles takes place over a memoryless
binary-input output-symmetric (MBIOS) channel, and the analysis
refers to the asymptotic case where we let the block length tend
to infinity. For the simplification of the numerical optimization,
a suitable approximation of the number of iterations was used in
\cite{ArdakaniSFK_Allerton05} to formulate this joint optimization
as a convex optimization problem. Due to the efficient tools which
currently exist for a numerical solution of convex optimization
problems, this approach suggests an engineering tool for the
design of good LDPC ensembles which possess an attractive tradeoff
between the decoding complexity and the asymptotic gap to capacity
(where the block length of these code ensembles is large enough).
This numerical approach however is not amenable for drawing
rigorous theoretical conclusions on the tradeoff between the
number of iterations and the performance of the code ensembles. A
different numerical approach for approximating the number of
iterations for LDPC ensembles operating over the BEC is addressed
in \cite{Ma_Yang_ISIT2004}.

A different approach for characterizing the complexity of
iterative decoders was suggested by Khandekar and McEliece (see
\cite{Khandekar-isit01, Khandekar_thesis,
McEliece_ISIT01_plenary_talk}). Their questions and conjectures
were related to the tradeoff between the asymptotic achievable
rates and the complexity under iterative message-passing decoding;
they initiated a study of the encoding and decoding complexity of
graph-based codes in terms of the achievable gap (in rate) to
capacity. It was conjectured there that for a large class of
channels, if the design rate of a suitably designed ensemble forms
a fraction $1-\varepsilon$ of the channel capacity, then the
decoding complexity scales like $\frac{1}{\varepsilon}
\ln\frac{1}{\varepsilon}$. The logarithmic term in this expression
was attributed to the graphical complexity (i.e., the decoding
complexity per iteration), and the number of iterations was
conjectured to scale like $\frac{1}{\varepsilon}$. There is one
exception: For the BEC, the complexity under the iterative
message-passing decoding algorithm behaves like $\ln
\frac{1}{\varepsilon}$ (see \cite{LubyMSS_IT01},
\cite{Sason-it03}, \cite{Sason-it04} and
\cite{Shokrollahi-IMA2000}). This is true since the absolute
reliability provided by the BEC allows every edge in the graph to
be used only once during the iterative decoding. Hence, for the
BEC, the number of iterations performed by the decoder serves
mainly to measure the delay in the decoding process, while the
decoding complexity is closely related to the complexity of the
Tanner graph which is chosen to represent the code. The graphical
complexity required for LDPC and systematic irregular
repeat-accumulate (IRA) code ensembles to achieve a fraction
$1-\varepsilon$ of the capacity of a BEC under iterative decoding
was studied in \cite{Sason-it03} and \cite{Sason-it04}. It was
shown in these papers that the graphical complexity of these
ensembles must scale at least like $\ln\frac{1}{\varepsilon}$;
moreover, some explicit constructions were shown to approach the
channel capacity with such a scaling of the graphical complexity.
An additional degree of freedom which is obtained by introducing
state nodes in the graph (e.g., punctured bits) was exploited in
\cite{PfisterSU_IT05} and \cite{Pfister_Sason_IT07} to construct
capacity-achieving ensembles of graph-based codes which achieve an
improved tradeoff between complexity and achievable rates.
Surprisingly, these capacity-achieving ensembles under iterative
decoding were demonstrated to maintain a {\em bounded graphical
complexity} regardless of the erasure probability of the BEC. A
similar result of a bounded graphical complexity for
capacity-achieving ensembles over the BEC was also obtained in
\cite{Hsu_IT_submitted}.

This paper provides simple lower bounds on the number of
iterations which is required for successful message-passing
decoding of graph-based code ensembles. The transmission of these
ensembles is assumed to take place over the BEC, and the bounds
refer to the asymptotic case where the block length tends to
infinity. The simplicity of the bounds derived in this paper stems
from the fact that they are easily evaluated and are expressed in
terms of some basic parameters of the considered ensemble; these
include the fraction of degree-2 variable nodes, the target bit
erasure probability and the gap between the channel capacity and
the design rate of the ensemble. The bounds derived in this paper
demonstrate that the number of iterations which is required for
successful message-passing decoding scales at least like the
inverse of the gap (in rate) to capacity, provided that the
fraction of degree-2 variable nodes of these turbo-like ensembles
does not vanish (hence, the number of iterations becomes unbounded
as the gap to capacity vanishes). The behavior of these lower
bounds matches well with the experimental results and the
conjectures on the number of iterations and complexity, as
provided by Khandekar and McEliece (see \cite{Khandekar-isit01},
\cite{Khandekar_thesis} and \cite{McEliece_ISIT01_plenary_talk}).
Note that lower bounds on the number of iterations in terms of the
target bit erasure probability can be alternatively viewed as
lower bounds on the achievable bit erasure probability as a
function of the number of iterations performed by the decoder. As
a result of this, the simple bounds derived in this paper provide
some insight on the design of stopping criteria for iteratively
decoded ensembles over the BEC (for other stopping criteria see,
e.g., \cite{Turbo-JPL, Stopping_criteria_1999}).

This paper is structured as follows: Section~\ref{Section:
Preliminaries} presents some preliminary background, definitions
and notation, Section~\ref{Section: main results} introduces the
main results of this paper and discusses some of their
implications, the proofs of these statements and some further
discussions are provided in Section~\ref{Section: Proofs of Main
Results}. Finally, Section~\ref{Section: Summary and Conclusions}
summarizes this paper. Proofs of some technical statements are
relegated to the appendices.

\section{Preliminaries}
\label{Section: Preliminaries}

This section provides preliminary background and introduces
notation for the rest of this paper.
\subsection{Graphical Complexity of Codes Defined on Graphs}
\label{subsection: Graphical complexity}

As noted in Section~\ref{Section: Introduction}, the decoding
complexity of a graph-based code under iterative message-passing
decoding is closely related to its graphical complexity, which we
now define formally.
\begin{definition}[Graphical Complexity]
{\em Let $\mathcal{C}$ be a binary linear block code of length $n$
and rate $R$, and let $\mathcal{G}$ be an arbitrary representation
of $\mathcal{C}$ by a Tanner graph. Denote the number of edges in
$\mathcal{G}$ by $E$. The graphical complexity of $\mathcal{G}$ is
defined as the number of edges in $\mathcal{G}$ per information
bit of the code $\mathcal{C}$, i.e., $\Delta(\mathcal{G})
\triangleq \frac{E}{nR}$\,.}
\end{definition}
Note that the graphical complexity depends on the specific Tanner
graph which is used to represent the code. An analysis of the
graphical complexity for some families of graph-based codes is
provided in \cite{Hsu_IT_submitted, PfisterSU_IT05,
Pfister_Sason_IT07, Sason-it03, Sason-it04}.

\subsection{Accumulate-Repeat-Accumulate Codes}
\label{subsection: ARA codes}

Accumulate-repeat-accumulate (ARA) codes form an attractive coding
scheme of turbo-like codes due to the simplicity of their encoding
and decoding (where both scale linearly with the block length),
and due to their remarkable performance under iterative decoding
\cite{Abbasfar_Comm07}. By some suitable constructions of
puncturing patterns, ARA codes with small maximal node degree are
presented in \cite{Abbasfar_Comm07}; these codes perform very well
even for short to moderate block lengths, and they suggest
flexibility in the design of efficient rate-compatible codes
operating on the same ARA decoder.

Ensembles of irregular and systematic ARA codes, which
asymptotically achieve the capacity of the BEC with bounded
graphical complexity, are presented in \cite{Pfister_Sason_IT07}.
This bounded complexity result stays in contrast to LDPC
ensembles, which have been shown to require unbounded graphical
complexity in order to approach channel capacity, even under
maximum-likelihood decoding (see \cite{Sason-it03}). In this
section, we present ensembles of irregular and systematic ARA
codes, and give a short overview of their encoding and decoding
algorithms; this overview is required for the later discussion.
The material contained in this section is taken from
\cite[Section~II]{Pfister_Sason_IT07}, and is introduced here
briefly in order to make the paper self-contained.

From an encoding point of view, ARA codes are viewed as
interleaved and serially concatenated codes. The encoding of ARA
codes is done as follows: first, the information bits are
accumulated (i.e., differentially encoded), and then the bits are
repeated a varying number of times (by an irregular repetition
code) and interleaved. The interleaved bits are partitioned into
disjoint sets (whose size is not fixed in general), and the parity
of each set of bits is computed (i.e., the bits are passed through
an irregular single parity-check (SPC) code). Finally, the bits
are accumulated a second time. A codeword of systematic ARA codes
is composed of the information bits and the parity bits at the
output of the second accumulator.

Since the iterative decoding algorithm of ARA codes is performed
on the appropriate Tanner graph (see Fig.~\ref{Figure: ARA tanner
graph}), this leads one to view them as sparse-graph codes from a
decoding point of view.

\begin{figure}[hbt]
\begin{center}
\epsfig{file=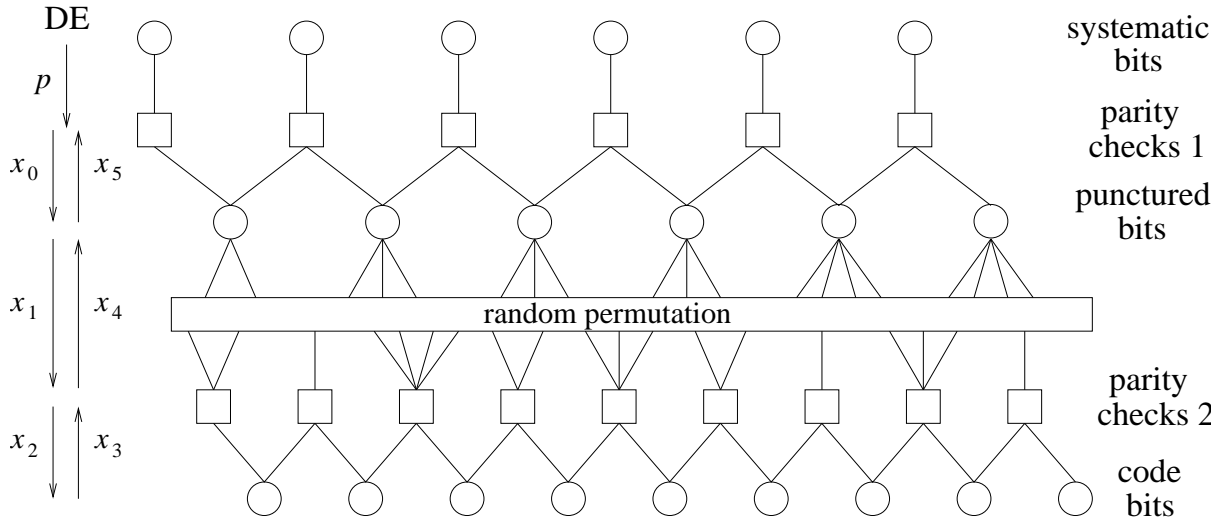,scale=0.7}
\end{center}
\caption{\label{Figure: ARA tanner graph} Tanner graph of an
irregular and systematic accumulate-repeat-accumulate code. This
figure is reproduced from~\cite{Pfister_Sason_IT07}.}
\end{figure}

Following the notation in \cite{Pfister_Sason_IT07}, we refer to
the three layers of bit nodes in the Tanner graphs as `systematic
bits' which form the systematic part of the codeword, `punctured
bits' which correspond to the output of the first accumulator and
are not a part of the transmitted codeword, and `code bits' which
correspond to the output of the second accumulator and form the
parity-bits of the codeword (see Fig.~\ref{Figure: ARA tanner
graph}). Denoting the block length of the code by $n$ and its
dimension by $k$, each codeword is composed of $k$ systematic bits
and $n-k$ code bits. The two layers of check nodes are referred to
as `parity-check~1' nodes and `parity-check~2' nodes, which
correspond to the first and the second accumulators of the
encoder, respectively. An ensemble of irregular ARA codes is
defined by the block length $n$ and the degree distributions of
the `punctured bit' and `parity-check~2' nodes. Following the
notation in \cite{Pfister_Sason_IT07}, the degree distribution of
the `punctured bit' nodes is given by the power series
\begin{equation}
L(x) \triangleq \sum_{i=1}^{\infty}L_i x^i \label{eq: ARA L}
\end{equation}
where $L_i$ designates the fraction of `punctured bit' nodes whose
degree is $i$. Similarly, the degree distribution of the
`parity-check~2' nodes is given by
\begin{equation}
R(x) \triangleq \sum_{i=1}^{\infty}R_ix^i \label{eq: ARA R}
\end{equation}
where $R_i$ designates the fraction of these nodes whose degree is
$i$. In both cases, degree of a node only refers to edges
connecting the `punctured bit' and the `parity-check~2' layers,
without the extra two edges which are connected to each of the
`punctured bit' nodes and `parity-check~2' nodes from the
accumulators (see Fig.~\ref{Figure: ARA tanner graph}).
Considering the distributions from the edge perspective, we let
\begin{equation}
\lambda(x) \triangleq \sum_{i=1}^{\infty}\lambda_i x^{i-1}, \quad
\rho(x) \triangleq \sum_{i=1}^{\infty}\rho_i x^{i-1} \label{eq: ARA
lambda and rho}
\end{equation}
designate the degree distributions from the edge perspective;
here, $\lambda_i$ ($\rho_i$) designates the fraction of edges
connecting `punctured bit' nodes to `parity-check~2' nodes which
are adjacent to `punctured bit' (`parity-check~2') nodes of degree
$i$. The design rate of a systematic ARA ensemble is given by
$R=\frac{a_{\text{R}}}{a_{\text{L}} + a_{\text{R}}} $ where
\begin{equation}
a_{\text{L}} \triangleq \sum_i i L_i = L'(1) =
\frac{1}{\dint_0^1\lambda(t)\mathrm{d}t} \; , \quad \quad
a_{\text{R}} \triangleq \sum_i i R_i = R'(1) =
\frac{1}{\dint_0^1\rho(t)\mathrm{d}t}\label{eq: ARA a_L and a_R}
\end{equation}
designate the average degrees of the `punctured bit' and
`parity-check~2' nodes, respectively.

Iterative decoding of ARA codes is performed by passing messages
on the edges of the Tanner graph in a layer-by-layer approach.
Each decoding iteration starts with messages for the `systematic
bit' nodes to the `parity-check~1' nodes, the latter nodes then
use this information to calculate new messages to the `punctured
bit' nodes and so the information passes through layers down the
graph and back up until the iteration ends with messages from the
`punctured bit' nodes to the `parity-check~1' nodes. The final
phase of messages from the `parity-check~1' nodes to the
`systematic bit' nodes is omitted since the latter nodes are of
degree one and so the outgoing message is not changed by incoming
information. Assume that the code is transmitted over a BEC with
erasure probability $p$. Since the systematic bits receive input
from the channel, the probability of erasure in messages from the
`systematic bit' nodes to the `parity-check~1' nodes is equal to
$p$ throughout the decoding process. For other messages, we denote
by $x_i^{(l)}$ where $i=0,1,\ldots,5$ the probability of erasure
of the different message types at decoding iteration number $l$
(where we start counting at zero). The variable $x_0^{(l)}$
corresponds to the probability of erasure in message from the
`parity-check~1' nodes to the `punctured bit' nodes, $x_1^{(l)}$
tracks the erasure probability of messages from the `punctured
bit' nodes to the `parity-check~2' nodes and so on. The density
evolution (DE) equations for the decoder based on the Tanner graph
in Figure~\ref{Figure: ARA tanner graph} are given in
\cite{Pfister_Sason_IT07}, and we repeat them here:
\begin{eqnarray}
\label{eq: ARA DE equations}
x_0^{(l)} &=& 1-\left(1-x_5^{(l-1)}\right)(1-p)\nonumber\\
x_1^{(l)} &=& \left(x_0^{(l)}\right)^2\,
\lambda\left(x_4^{(l-1)}\right)\nonumber\\
x_2^{(l)} &=&
1-R\left(1-x_1^{(l)}\right)\left(1-x_3^{(l-1)}\right)\quad\quad
l=1,2,\ldots\\
x_3^{(l)} & = & p\,x_2^{(l)}\nonumber\\
x_4^{(l)} & = &
1-\left(1-x_3^{(l)}\right)^2\rho\left(1-x_1^{(l)}\right)\nonumber\\
x_5^{(l)} & = & x_0^{(l)}\,L\left(x_4^{(l)}\right).\nonumber
\end{eqnarray}
The stability condition for systematic ARA ensembles is derived in
\cite[Section~II.D]{Pfister_Sason_IT07} and states that the fixed
point $x_i^{(l)} = 0$ of the iterative decoding algorithm is stable
if and only if
\begin{equation}
p^2\,\lambda_2\;\left(\rho'(1) + \frac{2pR'(1)}{1-p}\right)\leq
1\,.
  \label{eq: stability condition for ARA original version}
\end{equation}

\subsection{Big-O notation}
The terms $O$, $\Omega$ and $\Theta$ are widely used in computer
science to describe asymptotic relationships between functions
(for formal definitions see e.g., \cite{Big-O notation}). In our
context, we refer to the gap (in rate) to capacity, denoted by
$\varepsilon$, and discuss in particular the case where $0 \leq
\varepsilon \ll 1$ (i.e., sequences of capacity-approaching
ensembles). Accordingly, we define
\begin{itemize}
\item $f(\varepsilon) = O\bigl(g(\varepsilon)\bigr)$ means that there are
positive constants $c$ and $\delta$, such that $0 \leq
f(\varepsilon) \leq c \; g(\varepsilon)$ for all $0 \leq
\varepsilon \leq \delta$.
\item $f(\varepsilon) = \Omega\bigl(g(\varepsilon)\bigr)$ means that there are positive constants
$c$ and $\delta$, such that $0 \leq c \; g(\varepsilon) \leq
f(\varepsilon)$ for all $0 \leq \varepsilon \leq \delta$.
\item $f(\varepsilon) = \Theta\bigl(g(\varepsilon)\bigr)$ means that there are positive constants
$c_1$, $c_2$ and $\delta$, such that $0 \leq c_1 \; g(\varepsilon)
\leq f(\varepsilon)\leq c_2 \; g(\varepsilon)$ for all $0 \leq
\varepsilon \leq \delta$.
\end{itemize}
Note that for all the above definitions, the values of $c$, $c_1$,
$c_2$ and $\delta$ must be fixed for the function $f$ and should
not depend on~$\varepsilon$.

\section{Main Results}
\label{Section: main results}  In this section, we present lower
bounds on the required number of iterations used by a
message-passing decoder for code ensembles defined on graphs. The
communication is assumed to take place over a BEC, and we consider
the asymptotic case where the block length of these code ensembles
tends to infinity.
\begin{definition}
{\em Let $\big\{\mathcal{C}_m \big\}_{m\in\naturals}$ be a
sequence of code ensembles. Assume a common block length ($n_m$)
of the codes in $\mathcal{C}_m$ which tends to infinity as $m$
grows. Let the transmission of this sequence take place over a BEC
with capacity $C$. The sequence $\big\{\mathcal{C}_m \big\}$ is
said to {\em achieve a fraction} $1-\varepsilon$ {\em of the
channel capacity under some given decoding algorithm} if the
asymptotic rate of the codes in $\mathcal{C}_m$ satisfies $R\geq
(1-\varepsilon)C$ and the achievable bit erasure probability under
the considered algorithm vanishes as $m$ becomes large.}
\end{definition}
In the continuation, we consider a standard message-passing
decoder for the BEC, and address the number of iterations which is
required in terms of the achievable fraction of the channel
capacity under this decoding algorithm.

\begin{theorem}{\bf[Lower bound on the number of iterations
for LDPC ensembles transmitted over the BEC]} {\em Let
$\big\{(n_m, \lambda, \rho)\big\}_{m\in\naturals}$ be a sequence
of LDPC ensembles whose transmission takes place over a BEC with
erasure probability $p$. Assume that this sequence achieves a
fraction $1-\varepsilon$ of the channel capacity under
message-passing decoding. Let $L_2 = L_2(\varepsilon)$ be the
fraction of variable nodes of degree~2 for this sequence. In the
asymptotic case where the block length tends to infinity, let $l =
l(\varepsilon, p, P_{\mathrm{b}})$ denote the number of iterations
which is required to achieve an average bit erasure probability
$P_{\text{b}}$ over the ensemble. Under the mild condition that
$P_{\text{b}}<p\,L_2(\varepsilon)$, the required number of
iterations satisfies the lower bound
\begin{equation}
l(\varepsilon, p, P_{\mathrm{b}})\geq \frac{2}{1-p}
\left(\sqrt{p\,L_2(\varepsilon)} -\sqrt{P_{\text{b}}}\right)^2 \,
\frac{1}{\varepsilon}\,. \label{eq: lower bound on number of
iterations}
\end{equation}}
\label{theorem: number of iterations}
\end{theorem}

\begin{corollary}
{\em Under the assumptions of Theorem~\ref{theorem: number of
iterations}, if the fraction of degree-2 variable nodes stays
strictly positive as the gap (in rate) to capacity vanishes, i.e.,
if
\begin{equation*}
\lim_{\varepsilon \rightarrow 0} L_2(\varepsilon) > 0
\end{equation*}
then the number of iterations which is required in order to achieve
an average bit erasure probability
$P_{\text{b}}<p\,L_2(\varepsilon)$ under iterative message-passing
decoding scales at least like the inverse of this gap to capacity,
i.e.,}
\begin{equation*}
l(\varepsilon, p, P_{\mathrm{b}}) =
\Omega\left(\frac{1}{\varepsilon}\right).
\end{equation*}
\label{corollary: number of iterations}
\end{corollary}

\begin{discussion} {\bf[Effect of messages' scheduling on the number of iterations]}
{\em The lower bound on the number of iterations as provided in
Theorem~\ref{theorem: number of iterations} refers to the {\em
flooding schedule} where in each iteration, all the variable nodes
and subsequently all the parity-check nodes send messages to their
neighbors. Though it is the commonly used scheduling used by
iterative message-passing decoding algorithms, an alternative
scheduling of the messages may provide a faster convergence rate
for the iterative decoder. As an example, \cite{LDPC_scheduling}
considers the convergence rate of a {\em serial scheduling} where
instead of sending all the messages from the variable nodes to
parity-check nodes and then all the messages from check nodes to
variable nodes, as done in the flooding schedule, these two phases
are interleaved. Based on the density evolution analysis which
applies to the asymptotic case of an infinite block length, it is
demonstrated in \cite{LDPC_scheduling} that under some
assumptions, the required number of iterations for LDPC decoding
over the BEC with serial scheduling is reduced by a factor of two
(as compared to the flooding scheduling). It is noted that the
main result of Theorem~\ref{theorem: number of iterations} is the
introduction of a rigorous and simple lower bound on the number of
iterations for LDPC ensembles which scales like the reciprocal of
the gap between the channel capacity and the design rate of the
ensemble. Though such a scaling of this bound is proved for the
commonly used approach of flooding scheduling, it is likely to
hold also for other efficient approaches of scheduling. It is also
noted that this asymptotic scaling of the lower bound on the
number of iterations supports the conjecture of Khandekar and
McEliece \cite{Khandekar-isit01}.} \label{scheduling}
\end{discussion}

\begin{discussion} {\bf[On the dependence of the bounds on the
fraction of degree-2 variable nodes]} {\em  The lower bound on the
number of iterations in Theorem~\ref{theorem: number of
iterations} becomes trivial when the fraction of variable nodes of
degree~2 vanishes. Let us focus our attention on sequences of
ensembles which approach the channel capacity under iterative
message-passing decoding (i.e., $\varepsilon\rightarrow0$). For
the BEC, several such sequences have been constructed (see
e.g.~\cite{LubyMSS_IT01, Shokrollahi-IMA2000}). Asymptotically, as
the gap to capacity vanishes, all of these sequences known to date
satisfy the stability condition with equality; this property is
known as the flatness condition \cite{Shokrollahi-IMA2000}. In
\cite[Lemma~5]{Sason-submitted07}, the asymptotic fraction of
degree~2 variable nodes for capacity-approaching sequences of LDPC
ensembles over the BEC is calculated. This lemma states that for
such sequences which satisfy the following two conditions as the
gap to capacity vanishes:
\begin{itemize}
\item The stability condition is satisfied with equality (i.e.,
the flatness condition holds)
\item The limit of the ratio between the standard deviation and the
expectation of the right degree exists and is finite
\end{itemize}
then the asymptotic fraction of degree--2 variable nodes does not
vanish. In fact, for various sequences of capacity approaching
LDPC ensembles known to date (see \cite{LubyMSS_IT01, Oswald-it02,
Shokrollahi-IMA2000}), the ratio between the standard deviation
and the expectation of the right degree-distribution tends to
zero; in this case, \cite[Lemma~5]{Sason-submitted07} implies that
the fraction of degree-2 variable nodes tends to $\frac{1}{2}$
irrespectively of the erasure probability of the BEC, as can be
verified directly for these code ensembles.} \label{discussion:
L_2 for capacity approaching LDPC ensembles}
\end{discussion}

\begin{discussion} {\bf[Concentration of the lower bound]}
{\em Theorem~\ref{theorem: number of iterations} applies to the
required number of iterations for achieving an average bit erasure
probability $P_{\text{b}}$ where this average is taken over the LDPC
ensemble whose block length tends to infinity. Although we consider
an expectation over the LDPC ensemble, note that $l$ is
deterministic as it is the smallest integer for which the average
bit erasure probability does not exceed a fixed value. As shown in
the proof (see Section~\ref{Section: Proofs of Main Results}), the
derivation of this lower bound relies on the density evolution
technique which addresses the average performance of the ensemble.
Based on concentration inequalities, it is proved that the
performance of individual codes from the ensemble concentrates
around the average performance over the ensemble as we let the block
length tend to infinity \cite[Appendix~C]{Richardson_Urbanke_book}.
In light of this concentration result and the use of density
evolution in Section~\ref{Section: Proofs of Main Results} (which
applies to the case of an infinite block length), it follows that
the lower bound on the number of iterations in Theorem~\ref{theorem:
number of iterations} is valid with probability~1 for individual
codes from the ensemble. This also holds for the ensembles of codes
defined on graphs considered in Theorems~\ref{theorem: ARA number of
iterations} and~\ref{theorem: IRA number of iterations}.}
\label{discussion: concentration}
\end{discussion}

\begin{discussion} {\bf[On the number of required iterations for showing
a mild improvement in the erasure probability during the iterative
process]} {\em Note that for capacity-approaching LDPC ensembles,
the lower bound on the number of iterations tells us that even for
successfully starting the iteration process and reducing the bit
erasure probability by a factor which is below the fraction of
degree-2 variable nodes, the required number of iterations already
scales like $\frac{1}{\varepsilon}$. This is also the behavior of
the lower bound on the number of iterations even when the bit
erasure probability should be made arbitrarily small; this lower
bound therefore indicates that for capacity-approaching LDPC
ensembles, a significant number of the iterations is performed for
the starting process of the iterative decoding where the bit
erasure probability is merely reduced by a factor of $\frac{1}{2}$
as compared to the erasure probability of the channel (see
Discussion~\ref{discussion: L_2 for capacity approaching LDPC
ensembles} as a justification for the one-half factor). This
conclusion is also well interpreted by the area theorem and the
asymptotic behavior of the two EXIT curves (for the variable nodes
and the parity-check nodes) in the limit where $\varepsilon
\rightarrow 0$; as the gap to capacity vanishes, both curves tend
to be a step function jumping from $0$ to $1$ at the origin, so
the iterations progress very slowly at the initial stages of the
decoding process.}
\end{discussion}

In the asymptotic case where we let the block length tend to
infinity and the transmission takes place over the BEC, suitable
constructions of capacity-achieving systematic ARA ensembles
enable a fundamentally improved tradeoff between their graphical
complexity and their achievable gap (in rate) to capacity under
iterative decoding (see \cite{Pfister_Sason_IT07}). The graphical
complexity of these systematic ARA ensembles remains bounded (and
quite small) as the gap to capacity for these ensembles vanishes
under iterative decoding; this stays in contrast to un-punctured
LDPC code ensembles \cite{Sason-it03} and {\em systematic}
irregular repeat-accumulate (IRA) ensembles \cite{Sason-it04}
whose graphical complexity necessarily becomes unbounded as the
gap to capacity vanishes (see \cite[Table~I]{Pfister_Sason_IT07}).
This observation raises the question whether the number of
iterations which is required to achieve a desired bit erasure
probability under iterative decoding, can be reduced by using
systematic ARA ensembles. The following theorem provides a lower
bound on the number of iterations required to achieve a desired
bit erasure probability under message-passing decoding; it shows
that similarly to the parallel result for LDPC ensembles (see
Theorem~\ref{theorem: number of iterations}), the required number
of iterations for systematic ARA codes scales at least like the
inverse of the gap to capacity.

\begin{theorem} {\bf[Lower bound on the number of iterations for systematic ARA
ensembles transmitted over the BEC]} {\em Let $\big\{(n_m,
\lambda, \rho)\big\}_{m\in\naturals}$ be a sequence of systematic
ARA ensembles whose transmission takes place over a BEC with
erasure probability $p$. Assume that this sequence achieves a
fraction $1-\varepsilon$ of the channel capacity under
message-passing decoding. Let $L_2 = L_2(\varepsilon)$ be the
fraction of `punctured bit' nodes of degree~2 for this sequence
(where the two edges related to the accumulator are not taken into
account). In the asymptotic case where the block length tends to
infinity, let $l = l(\varepsilon, p, P_{\mathrm{b}})$ designate
the required number of iterations to achieve an average bit
erasure probability $P_{\text{b}}$ of the systematic bits. Under
the mild condition that $1-\sqrt{1-\frac{P_{\text{b}}}{p}} <
p\,L_2(\varepsilon)$, the number of iterations satisfies the lower
bound
\begin{equation}
l(\varepsilon, p, P_{\mathrm{b}}) \geq 2p(1-\varepsilon)\,
  \left(\sqrt{p\,L_2(\varepsilon)}-\sqrt{1-\sqrt{1-
  \frac{P_{\text{b}}}{p}}}
  \right)^2 \; \frac{1}{\varepsilon} \label{eq: ARA lower bound on
number of iterations}\,.
\end{equation}
\label{theorem: ARA number of iterations}}
\end{theorem}

As noted in Section~\ref{subsection: ARA codes}, systematic ARA
codes can be viewed as serially concatenated codes where the
systematic bits are associated with the outer code. These codes
can be therefore decoded iteratively by using a turbo-like decoder
for interleaved and serially concatenated codes. The following
proposition states that the lower bound on the number of
iterations in Theorem~\ref{theorem: ARA number of iterations} is
also valid for such an iterative decoder.
\begin{proposition} {\bf[Lower bound on the number of
iterations for systematic ARA codes under turbo-like decoding]}
{\em Under the assumptions and notation of Theorem~\ref{theorem:
ARA number of iterations}, the lower bound on the number of
iterations in \eqref{eq: ARA lower bound on number of iterations}
is valid also when the decoding is performed by a turbo-like
decoder for uniformly interleaved and serially concatenated
codes.} \label{proposition: ARA number of iterations turbo}
\end{proposition}
The reader is referred to \ref{Appendix: Proof of proposition ARA
number of iterations turbo} for a detailed proof. The following
theorem which refers to irregular repeat-accumulate (IRA)
ensembles is proved in a conceptually similar way to the proof of
Theorem~\ref{theorem: ARA number of iterations}.

\begin{theorem} {\bf[Lower bound on the number of iterations for IRA
ensembles transmitted over the BEC]} {\em Let $\big\{(n_m,
\lambda, \rho)\big\}_{m\in\naturals}$ be a sequence of (systematic
or non-systematic) IRA ensembles whose transmission takes place
over a BEC with erasure probability $p$. Assume that this sequence
achieves a fraction $1-\varepsilon$ of the channel capacity under
message-passing decoding. Let $L_2 = L_2(\varepsilon)$ be the
fraction of `information bit' nodes of degree~2 for this sequence.
In the asymptotic case where the block length tends to infinity,
let $l = l(\varepsilon, p, P_{\mathrm{b}})$ designate the required
number of iterations to achieve an average bit erasure probability
$P_{\text{b}}$ of the information bits. For systematic codes, if
$P_{\text{b}} < p\,L_2(\varepsilon)$, then the number of
iterations satisfies the lower bound
\begin{equation}
l(\varepsilon, p, P_{\mathrm{b}}) \geq 2(1-\varepsilon)
  \left(\sqrt{p\,L_2(\varepsilon)}-\sqrt{P_{\text{b}}}\right)^2
  \;\frac{1}{\varepsilon} \label{eq: systematic IRA lower
bound on number of iterations}\,.
\end{equation}
For non-systematic codes, if $P_{\text{b}}<L_2(\varepsilon)$, then
\begin{equation}
l(\varepsilon, p, P_{\mathrm{b}}) \geq 2(1-\varepsilon)
  \left(\sqrt{L_2(\varepsilon)}-\sqrt{P_{\text{b}}}\right)^2
  \;\frac{1}{\varepsilon}. \label{eq: non-systematic IRA lower
bound on number of iterations}
\end{equation}
\label{theorem: IRA number of iterations}}
\end{theorem}

\section{Derivation of the Bounds on the Number of Iterations}
\label{Section: Proofs of Main Results}
\subsection{Proof of Theorem~\ref{theorem: number of iterations}}
\label{Subsection: Proof of Theorem: number of iterations} Let
$\big\{x^{(l)}\big\}_{l\in\naturals}$ designate the expected
fraction of erasures in messages from the variable nodes to the
check nodes at the $l$'th iteration of the message-passing
decoding algorithm (where we start counting at $l=0$). From
density evolution, in the asymptotic case where the block length
tends to infinity, $x^{(l)}$ is given by the recursive equation
\begin{equation}
x^{(l+1)} = p \;
\lambda\left(1-\rho\big(1-x^{(l)}\big)\right)\,,\quad
l\in\naturals \label{eq: recursive equation for x_l}
\end{equation}
with the initial condition
\begin{equation}
x^{(0)} = p \label{eq: initial condition for x_l}
\end{equation}
where $p$ designates the erasure probability of the BEC.
Considering a sequence of $\{(n_m, \lambda, \rho)\}$ LDPC
ensembles where we let the block length $n_m$ tend to infinity,
the average bit erasure probability after the $l$'th iteration is
given by
\begin{equation}
P_{\text{b}}^{(l)} = p \; L\big(1-\rho(1-x^{(l)})\big)
  \label{eq: bit erasure probability at l'th iteration}
\end{equation}
where $L$ designates the common left degree distribution of the
ensembles from the node perspective. Since the function $f(x) =
p\,\lambda\big(1-\rho(1-x)\big)$ is monotonically increasing,
Eqs.~\eqref{eq: recursive equation for x_l}--\eqref{eq: bit
erasure probability at l'th iteration} imply that an average bit
erasure probability of $P_{\text{b}}$ is attainable under
iterative message-passing decoding if and only if
\begin{equation}
p \; \lambda\big(1-\rho(1-x)\big) < x \,,\quad \forall x\in(x^*,p]
\label{eq: condition to achieve P_b}
\end{equation}
where $x^*$ is the unique solution of
\begin{equation*}
P_{\text{b}} = p\; L\big(1-\rho(1-x^*)\big)\,.
\end{equation*}
Let us define the functions
\begin{equation}
  c(x) \triangleq 1-\rho(1-x), \quad \quad v(x) =
  \left\{\begin{tabular}{ll}
  $\lambda^{-1}\left(\frac{x}{p}\right)$ & $0\leq x\leq p$\\
  $1$ & $p < x \leq 1$\end{tabular}\right..
  \label{eq: definition of c(x) and v(x)}
\end{equation}
From the condition in \eqref{eq: condition to achieve P_b}, an
average bit erasure probability of $P_{\text{b}}$ is attained if
and only if $c(x)<v(x)$ for all $x\in(x^*,p]$. Since we assume
that vanishing bit erasure probability is achievable under
message-passing decoding, it follows that $c(x)<v(x)$ for all
$x\in(0,p]$.
\begin{figure}[here!]
\begin{center}
\epsfig{file=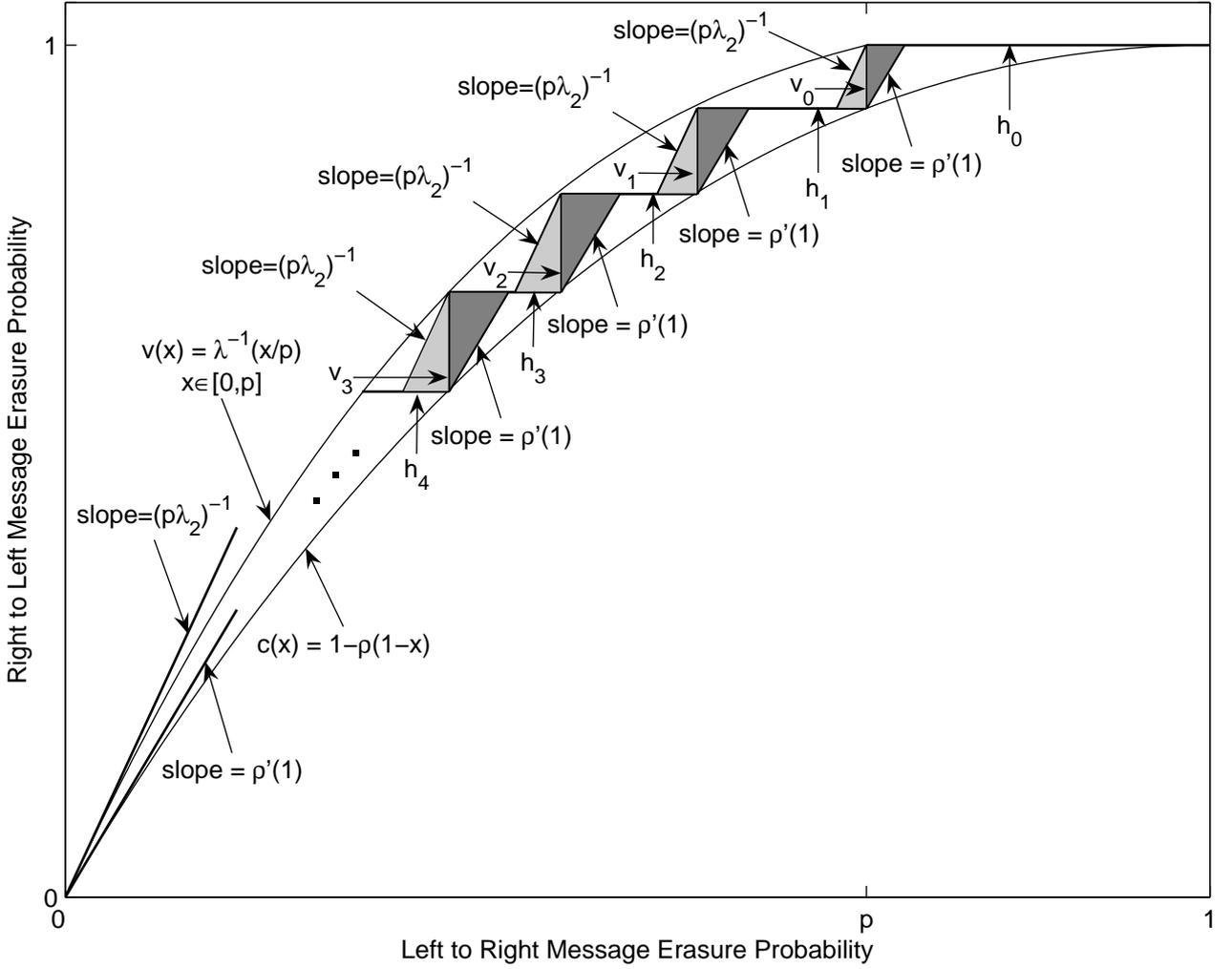,scale=1}
\end{center}
\caption{\label{Figure: Fig for proof of Thm 1} Plot of the
functions $c(x)$ and $v(x)$ for an ensemble of LDPC codes which
achieves vanishing bit erasure probability under iterative
message-passing decoding when communicated over a BEC whose
erasure probability is equal to $p$. The horizontal and vertical
lines track the evolution of the expected fraction of erasure
messages from the variable nodes to the check nodes at each
iteration of the message-passing decoding algorithm.}
\end{figure}
Figure~\ref{Figure: Fig for proof of Thm 1} shows a plot of the
functions $c(x)$ and $v(x)$ for an ensemble of LDPC codes which
achieves vanishing bit erasure probability under iterative
decoding as the block length tends to infinity. The horizontal and
vertical lines, labeled $\big\{h_l\big\}_{l\in\naturals}$ and
$\big\{v_l\big\}_{l\in\naturals}$, respectively, are used to track
the expected fraction of erased messages from the variable nodes
to the parity-check nodes at each iteration of the message-passing
decoding algorithm. From \eqref{eq: recursive equation for x_l}
and \eqref{eq: initial condition for x_l}, the expected fraction
of erased left to right messages in the $l$'th decoding iteration
(where we start counting at zero) is equal to the $x$ value at the
left tip of the horizontal line $h_l$. The right-angled triangles
shaded in gray will be used later in the proof.

The first step in the proof of Theorem~\ref{theorem: number of
iterations} is calculating the area bounded by the curves $c(x)$
and $v(x)$. This is done in the following lemma which is based on
the area theorem for the BEC \cite{Ashikhmin}.
\begin{lemma}{\em \label{lemma: area between c(x) and v(x)}
\begin{equation}
\int_{0}^{1}\big(v(x)-c(x)\big)\mathrm{d}x =
\frac{C-R}{a_{\text{L}}}
  \label{eq: area between c(x) and v(x)}
\end{equation}
where $C=1-p$ is the capacity of the BEC, $R$ is the design rate of
the ensemble, and $a_{\text{L}}$ is the average left degree of the
ensemble.}
\end{lemma}
\begin{proof}
The definitions of the functions $v$ and $c$ in \eqref{eq:
definition of c(x) and v(x)} imply that
\begin{eqnarray*}
\int_{0}^{1}\big(v(x)-c(x)\big)\mathrm{d}x & = &
\int_{0}^{p}\lambda^{-1}\left(\frac{x}{p}\right)\mathrm{d}x +
\int_{p}^{1}1\,\mathrm{d}x
 - \int_{0}^{1}c(x)\mathrm{d}x\\
 &=& p \int_{0}^{1}\lambda^{-1}(s)\mathrm{d}s + 1-p -
 \int_{0}^{1} \big(1-\rho(1-x)\big)\mathrm{d}x\\
&\stackrel{(a)}{=}&
p\left(1-\int_{0}^{1}\lambda(x)\mathrm{d}x\right) + 1- p - 1 +
\int_{0}^{1}\rho(x)\mathrm{d}x \\
&=& \int_{0}^{1}\rho(x)\mathrm{d}x - p\int_{0}^{1}\lambda(x)\mathrm{d}x\\
&=&\underbrace{\int_{0}^{1}\lambda(x)\mathrm{d}x}_{\frac{1}{a_{\text{L}}}}
\left(\underbrace{\frac{\int_{0}^{1}\rho(x)\mathrm{d}x}{\int_{0}^{1}\lambda(x)\mathrm{d}x}}_{1-R}
-\underbrace{p}_{1-C}\right)\\ &=&\frac{C-R}{a_{\text{L}}}
  \end{eqnarray*}
where $(a)$ follows by substituting $x = \lambda(s)$ and applying
integration by parts.
\end{proof}

Let us consider the two sets of right-angled triangles shown in two shades of gray in
Figure~\ref{Figure: Fig for proof of Thm 1}. The set of triangles which are shaded in dark gray are
defined so that one of the legs of triangle number $i$ (counting
from right to left and starting at zero) is the vertical line $v_i$,
and the slope of the hypotenuse is equal to $c'(0) = \rho'(1)$.
Since $c(x)$ is concave for all $x\in[0,1]$, these triangles are
guaranteed to be above the curve of the function $c$.
Since the slope of the hypotenuse is $\rho'(1)$, the area of the
$i$'th triangle in this set is
\begin{equation}
A_i = \frac{1}{2}\;|v_i|\;\left(\frac{|v_i|}{\rho'(1)}\right) =
\frac{|v_i|^2}{2\rho'(1)} \label{eq: area of dark triangle}
\end{equation}
where $|v_i|$ is the length of $v_i$. We now turn to consider the
second set of triangles, which are shaded in light gray. Note that
the function $\lambda(x)$ is monotonically increasing and convex
in $[0,1]$ and also that $\lambda(0)=0$ and $\lambda(1)=1$. This
implies that $\lambda^{-1}$ is concave in $[0,1]$ and therefore
$v(x)$ is concave in $[0,p]$. The triangles shaded in light gray
are defined so that one of the legs of triangle number $i$ (again,
counting from the right and starting at zero) is the vertical line
$v_i$ and the slope of the hypotenuse is given by
\begin{equation*}
v'(0) = \frac{1}{p}\left(\lambda^{-1}\right)'(0) = \frac{1}{p\lambda'(0)} = \frac{1}{p\lambda_2}
\end{equation*}
where the second equality follows since $\lambda(0)=0$. The
concavity of $v(x)$ in $[0,p]$ guarantees that these triangles are
below the curve of the of function $v$. The area of the $i$'th
triangle in this second set of triangles is given by
\begin{equation}
B_i = \frac{1}{2}\;|v_i|\;\left(|v_i|\,p\lambda_2\right) =
\frac{p\lambda_2\,|v_i|^2}{2}\,. \label{eq: area of light
triangle}
\end{equation}
Since $v(x)$ is monotonically increasing with $x$, the dark-shaded
triangles lie below the curve of the function $v$. Similarly, the
monotonicity of $c(x)$ implies that the light-shaded triangles are
above the curve of the function $c$. Hence, both sets of triangles
form a  subset of the domain bounded by the curves of $c(x)$ and
$v(x)$. By their definitions, the $i$'th dark triangle is on the
right of $v_i$, and the $i$'th light triangle lies to the left of
$v_i$; therefore, the triangles do not overlap. Combining \eqref{eq:
area of dark triangle}, \eqref{eq: area of light triangle} and the
fact that the triangles do not overlap, and applying
Lemma~\ref{lemma: area between c(x) and v(x)}, we get
\begin{eqnarray}
\frac{C-R}{a_{\text{L}}} &=& \int_0^1 \bigl(v(x)-c(x) \bigr) \mathrm{d}x\nonumber\\
&\geq&\sum_{i=0}^\infty \left(A_i+B_i\right)\nonumber\\
&\geq& \frac{1}{2}\,\left(\frac{1}{\rho'(1)}+p\lambda_2\right)\,\sum_{i=0}^{l-1}|v_i|^2
 \label{eq: area of triangles}
\end{eqnarray}
where $l$ is an arbitrary natural number. Since we assume that the
bit erasure probability vanishes under iterative message-passing
decoding, the stability condition implies that
\begin{equation}
\frac{1}{\rho'(1)} \geq p\,\lambda_2\,.
  \label{eq: stability condition for the BEC}
\end{equation}
Substituting \eqref{eq: stability condition for the BEC} and $R =
(1-\varepsilon)C$ in \eqref{eq: area of triangles} gives
\begin{equation}
    C\varepsilon \geq
    a_{\text{L}}\,p\lambda_2\,\sum_{i=0}^{l-1}|v_i|^2 .
  \label{eq: lower bound on epsilonC}
\end{equation}
The definition of $h_l$ and $v_l$ in Figure~\ref{Figure: Fig for
proof of Thm 1} implies that for an arbitrary iteration $l$
\begin{equation*}
1-\rho(1-x^{(l)}) = c(x^{(l)}) = 1-\sum_{i=0}^l |v_i|\,.
\end{equation*}
Substituting the last equality in \eqref{eq: bit erasure
probability at l'th iteration} yields that the average bit erasure
probability after iteration number $l-1$ can be expressed as
\begin{equation}
P_{\text{b}}^{(l-1)} = p\; L\left(1-\sum_{i=0}^{l-1}
|v_i|\right)\,.
  \label{eq: bit erasure probability at l'th iteration as function of
  v_i}
\end{equation}
Let $l$ designate the number of iterations required to achieve an
average bit erasure probability $P_{\text{b}}$ over the ensemble
(where we let the block length tend to infinity), i.e., $l$ is the
smallest integer which satisfies $P_{\text{b}}^{(l-1)}\leq
P_{\text{b}}$ since we start counting at $l=0$. Although we
consider an expectation over the LDPC ensemble, note that $l$ is
deterministic as it is the smallest integer for which the average
bit erasure probability does not exceed $P_{\text{b}}$. Since $L$
is monotonically increasing, \eqref{eq: bit erasure probability at
l'th iteration as function of v_i} provides a lower bound on
$\sum_{i=0}^{l-1} |v_i|$ of the form
\begin{equation}
\sum_{i=0}^{l-1} |v_i| \geq
1-L^{-1}\left(\frac{P_{\text{b}}}{p}\right)\,.
  \label{eq: lower bound on sum of v_i}
\end{equation}
From the Cauchy-Schwartz inequality, we get
\begin{equation}
\left(\sum_{i=0}^{l-1} |v_i|\right)^2 \leq
\sum_{i=0}^{l-1}1\,\sum_{i=0}^{l-1} |v_i|^2 = l\sum_{i=0}^{l-1}
|v_i|^2. \label{eq: Cauchy-Schwarz inequality for sum of v_i}
\end{equation}
Combining the above inequality with \eqref{eq: lower bound on
epsilonC} and \eqref{eq: lower bound on sum of v_i} gives the
inequality
\begin{equation*}
C\varepsilon \geq
\frac{a_{\text{L}}\, p\lambda_2 \left(1-L^{-1}\Big(\frac{P_{\text{b}}}{p}\Big)\right)^2}{l}
\end{equation*}
which provides the following lower bound on the number of iterations
$l$:
\begin{equation}
l \geq
\frac{a_{\text{L}}\, p\lambda_2 \left(1-L^{-1}\Big(\frac{P_{\text{b}}}{p}\Big)\right)^2}
{(1-p)\varepsilon}.
  \label{eq: first lower bound on number of iterations}
\end{equation}
To continue the proof, we derive a lower bound on $1-L^{-1}(x)$ for $x\in(0,1)$. Since
the fraction of variable nodes of degree $i$ is non-negative for all
$i=2,3,\ldots$, we have
\begin{equation*}
  L(x) = \sum_{i}L_i x^i \geq L_2 x^2, \quad x \geq 0.
\end{equation*}
Substituting $t=L(x)$ gives
\begin{equation*}
  t \geq L_2\cdot\big(L^{-1}(t)\big)^2,\quad \forall
  t\in(0,1)
\end{equation*}
which is transformed into the following lower bound on
$1-L^{-1}(x)$:
\begin{equation}
1-L^{-1}(x) \geq 1-\sqrt{\frac{x}{L_2}}\,,\quad \forall x\in(0,1)\,.
\label{eq: lower bound on 1-L^-1}
\end{equation}
Under the assumption $\frac{P_{\text{b}}}{p}<L_2$, substituting
\eqref{eq: lower bound on 1-L^-1} in \eqref{eq: first lower bound
on number of iterations} gives
\begin{eqnarray}
l &\geq& \frac{a_{\text{L}}\, p\lambda_2 \left(\sqrt{L_2} -
\sqrt{\frac{P_{\text{b}}}{p}}\right)^2}
{L_2\,(1-p)\varepsilon}\nonumber\\
&=&\frac{a_{\text{L}}\,\lambda_2 \left(\sqrt{p\,L_2} -
\sqrt{P_{\text{b}}}\right)^2} {L_2\,(1-p)\varepsilon}\,.
  \label{eq: second lower bound on number of iterations}
\end{eqnarray}
The lower bound in \eqref{eq: lower bound on number of iterations}
is obtained by substituting the equality $ L_2 = \frac{\lambda_2
\,a_{\text{L}}}{2}$ into \eqref{eq: second lower bound on number
of iterations}.

Taking the limit where the average bit erasure probability tends
to zero on both sides of \eqref{eq: lower bound on number of
iterations} gives the following lower bound on the number of
iterations:
\begin{equation*}
l(\varepsilon, p, P_{\text{b}}\rightarrow 0) \geq \frac{2p}{1-p}
\frac{L_2(\varepsilon)}{\varepsilon}\,.
\end{equation*}

\subsection{Proof of Theorem~\ref{theorem: ARA number of iterations}}
We begin the proof by considering the expected fraction of erasure
messages from the `punctured bit' nodes to the `parity-check~2'
nodes (see Fig.~\ref{Figure: ARA tanner graph}). The following
lemma provides a lower bound on the expected fraction of erasures
in the $l$'th decoding iteration in terms of this expected
fraction at the preceding iteration.
\begin{lemma}{\em
Let $(n, \lambda, \rho)$ be an ensemble of systematic ARA codes
whose transmission takes place over a BEC with erasure probability
$p$. Then, in the limit where the block length tends to infinity,
the expected fraction of erasure messages from the `punctured bit'
nodes to the `parity-check~2' nodes at the $l$'th iteration
satisfies
\begin{equation}
  x_1^{(l)} \geq \widetilde{\lambda}\left(1-\widetilde{\rho}\bigl(1-x_1^{(l-1)}\bigr)
  \right),\quad l=1,2,\ldots
  \label{eq: x_1 as a function of previous x_1}
\end{equation}
where the tilted degree distributions $\widetilde{\lambda}$ and
$\widetilde{\rho}$ are given as follows (see
\cite{Pfister_Sason_IT07}):
\begin{eqnarray}
&& \widetilde{\lambda}(x) \triangleq \left(\frac{p}{1-(1-p)L(x)}\right)^2\lambda(x)\ \label{eq: lambda tilde}\\
&& \widetilde{\rho}(x) \triangleq \left(\frac{1-p}{1-p
R(x)}\right)^2\rho(x) \label{eq: rho tilde}
\end{eqnarray}
and $L$ and $R$ designate the degree distributions of the ARA
ensemble from the node perspective.} \label{lemma: x_1 as a
function of previous x_1}
\end{lemma}
\begin{proof}
See Appendix~\ref{Appendix: Proof of lemma: x_1 as a function of
previous x_1}.
\end{proof}

From Fig.~\ref{Figure: ARA tanner graph}, it can be readily
verified that the probabilities $x_0$ and $x_1$ for erasure
messages at iteration no. zero are equal to~1, i.e.,
\begin{equation}
x_0^{(0)}=x_1^{(0)}=1. \label{eq: ARA initial condition for x_1}
\end{equation}
Let us look at the RHS of \eqref{eq: x_1 as a function of previous
x_1} as a function of $x$, and observe that it is monotonically
increasing over the interval $[0,1]$. Let us compare the
performance of a systematic ARA ensemble whose degree
distributions are $(\lambda, \rho)$ with an LDPC ensemble whose
degree distributions are given by $(\widetilde{\lambda},
\widetilde{\rho})$ (see \eqref{eq: lambda tilde} and \eqref{eq:
rho tilde}) under iterative message-passing decoding. Given the
initial condition $x_1^{(0)}=1$, the following conclusion is
obtained by recursively applying Lemma~\ref{lemma: x_1 as a
function of previous x_1}: For any iteration, the erasure
probability for messages delivered from `punctured bit' nodes to
`parity-check~2' nodes of the ARA ensemble (see Fig.~\ref{Figure:
ARA tanner graph}) is lower bounded by the erasure probability of
the left-to-right messages of the LDPC ensemble; this holds even
if the a-priori information from the BEC is not used by the
iterative decoder of the LDPC ensemble (note that the coefficient
of $\widetilde{\lambda}$ in the RHS of \eqref{eq: x_1 as a
function of previous x_1} is equal to one). Note that unless the
fraction of `parity-check~2' nodes of degree~1 is strictly
positive (i.e., $R_1 > 0$), the iterative decoding cannot be
initiated for both ensembles (unless some the values of some
'punctured bits' of the systematic ARA ensemble are known, as in
\cite{Pfister_Sason_IT07}). Hence, the comparison above between
the ARA and LDPC ensembles is of interest under the assumption
that $R_1 > 0$; this property is implied by the assumption of
vanishing bit erasure probability for the systematic ARA ensemble
under iterative message-passing decoding.

In \cite[Section~II.C.2]{Pfister_Sason_IT07}, a technique called
`graph reduction' is introduced. This technique transforms the
Tanner graph of a systematic ARA ensemble, transmitted over a BEC
whose erasure probability is $p$, into a Tanner graph of an
equivalent LDPC ensemble (where this equivalence holds in the
asymptotic case where the block length tends to infinity). The
variable and parity-check nodes of the equivalent LDPC code evolve
from the `punctured bit' and `parity-check 2' nodes of the ARA
ensemble, respectively, and their degree distributions (from the
edge perspective) are given by $\widetilde{\lambda}$ and
$\widetilde{\rho}$, respectively. It is also shown in
\cite{Pfister_Sason_IT07} that $\widetilde{\lambda}$ and
$\widetilde{\rho}$ are legitimate degree distribution functions,
i.e., all the derivatives at zero are non-negative and
$\widetilde{\lambda}(1) = \widetilde{\rho}(1) = 1$. As shown in
\cite[Eqs.~(9)--(12)]{Pfister_Sason_IT07}, the left and right
degree distributions of the equivalent LDPC ensemble from the node
perspective are given, respectively, by
\begin{equation}
  \widetilde{L}(x) = \frac{\dint_0^x \widetilde{\lambda}(t)\mathrm{d}t}{\dint_0^1
  \widetilde{\lambda}(t)\mathrm{d}t} = \frac{p\,L(x)}{1-(1-p)L(x)}
  \label{eq: L tilde}
\end{equation}
and
\begin{equation}
\widetilde{R}(x) = \frac{\dint_0^x
\widetilde{\rho}(t)\mathrm{d}t}{\dint_0^1
  \widetilde{\rho}(t)\mathrm{d}t} = \frac{(1-p)\,R(x)}{1-pR(x)} \; . \label{eq: R tilde}
\end{equation}

Let $P_{\text{b}}^{(l)}$ designate the average erasure probability
of the systematic bits after the $l$'th decoding iteration (where
we start counting at $l=0$). For LDPC ensembles, a simple
relationship between the erasure probability of the code bits and
the erasure probability of the left-to-right messages at the
$l$'th decoding iteration is given in \eqref{eq: bit erasure
probability at l'th iteration}. For systematic ARA ensembles, a
similar, though less direct, relationship exists between the
erasure probability of the systematic bits after the $l$'th
decoding iteration and $x_1^{(l)}$; this relationship is presented
in the following lemma.
\begin{lemma}
{\em Let $(n, \lambda, \rho)$ be an ensemble of systematic ARA
codes whose transmission takes place over a BEC with erasure
probability $p$. Then, in the asymptotic case where the block
length tends to infinity, the average erasure probability of the
systematic bits after the $l$'th decoding iteration,
$P_{\text{b}}^{(l)}$, satisfies the inequality
\begin{equation}
1-\sqrt{1-\frac{P_{\text{b}}^{(l)}}{p}} \geq
\widetilde{L}\left(1-\widetilde{\rho}\left(1-x_1^{(l)}\right)\right)
\label{eq: P_b,l for ARA}
\end{equation}
where $\widetilde{\rho}$ and $\widetilde{L}$ are defined in
\eqref{eq: rho tilde} and \eqref{eq: L tilde}, respectively
(similarly to their definitions in \cite{Pfister_Sason_IT07}).}
\label{lemma: P_b,l for ARA}
\end{lemma}
\begin{proof}
See Appendix~\ref{Appendix: Proof of lemma: P_b,l for ARA}.
\end{proof}
\begin{remark}
{\em We note that when $P_{\text{b}}^{(l)}$ is very small, the LHS
of \eqref{eq: P_b,l for ARA} satisfies
\begin{equation*}
1-\sqrt{1-\frac{P_{\text{b}}^{(l)}}{p}} \approx
\frac{P_{\text{b}}^{(l)}}{2p}\,,
\end{equation*}
so \eqref{eq: P_b,l for ARA} takes a similar form to \eqref{eq:
bit erasure probability at l'th iteration} which refers to the
erasure probability of LDPC ensembles.}
\end{remark}
Consider the number of iterations required for the message-passing
decoder, operating on the Tanner graphs of the systematic ARA
ensemble, to achieve a desired bit erasure probability
$P_{\text{b}}$. Combining Lemmas~\ref{lemma: x_1 as a function of
previous x_1} and~\ref{lemma: P_b,l for ARA}, and the initial
condition in \eqref{eq: ARA initial condition for x_1}, a lower
bound on this number of iterations can be deduced. More
explicitly, it is lower bounded by the number of iterations which
is required to achieve a bit erasure probability of
$1-\sqrt{1-\frac{P_{\text{b}}}{p}}$ for the LDPC ensemble whose
degree distributions are $(\widetilde{\lambda}, \widetilde{\rho})$
and where the erasure probability of the BEC is equal to~1. It is
therefore tempting to apply the lower bound on the number of
iterations in Theorem~\ref{theorem: number of iterations}, which
refers to LDPC ensembles, as a lower bound on the number of
iterations for the ARA ensemble. Unfortunately, the LDPC ensemble
with the tilted pair of degree distributions
$(\widetilde{\lambda},\widetilde{\rho})$ is transmitted over a BEC
whose erasure probability is 1, so the channel capacity is equal
to zero and the multiplicative gap to capacity is meaningless.
This prevents a direct use of Theorem~\ref{theorem: number of
iterations}; however, the continuation of the proof follows
similar lines in the proof of Theorem~\ref{theorem: number of
iterations}.

Let $x^*$ denote the unique solution in $[0,1]$ of the equation
\begin{equation}
1-\sqrt{1-\frac{P_{\text{b}}}{p}} =
\widetilde{L}\bigl(1-\widetilde{\rho}\left(1-x^*\right)\bigr).
\label{eq: equation for x*}
\end{equation}
From \eqref{eq: x_1 as a function of previous x_1}, \eqref{eq: ARA
initial condition for x_1} and \eqref{eq: P_b,l for ARA}, a
necessary condition for achieving a bit erasure probability
$P_{\text{b}}$ of the systematic bits is that
\begin{equation}
\widetilde{\lambda}\bigl(1-\widetilde{\rho}(1-x)\bigr)<x\, , \quad
\forall x\in(x^*,1]\,. \label{eq: condition to achieve P_b in ARA}
\end{equation}
In the limit where the fixed point of the iterative decoding
process is attained, the inequalities in \eqref{eq: x_1 as a
function of previous x_1}, \eqref{eq: ARA initial condition for
x_1} and \eqref{eq: P_b,l for ARA} are replaced by equalities;
hence, \eqref{eq: condition to achieve P_b in ARA} also forms a
sufficient condition. Analogously to the case of LDPC ensembles,
as in the proof of Theorem~\ref{theorem: number of iterations}, we
define the functions
\begin{equation}
  \widetilde{c}(x) = 1-\widetilde{\rho}(1-x)\quad\text{and}\quad v(x) =
  \widetilde{\lambda}^{-1}(x)\,.
\label{eq: tilde c and tilde v}
\end{equation}
Due to the monotonicity of $\widetilde{\lambda}$ in $[0,1]$, the
necessary and sufficient condition for attaining an erasure
probability $P_{\text{b}}$ of the systematic bits in \eqref{eq:
condition to achieve P_b in ARA} can be rewritten as
\begin{eqnarray*}
  \widetilde{c}(x) < \widetilde{v}(x)\, , \quad \forall x\in(x^*,1]\,.
\end{eqnarray*}
 Since we assume that the sequence of ensembles asymptotically achieves
vanishing bit erasure probability under message-passing decoding, it
follows that
\begin{eqnarray*}
  \widetilde{c}(x) < \widetilde{v}(x)\, , \quad \forall x\in(0,1]\,.
\end{eqnarray*}
The next step in the proof is calculating the area of the domain
bounded by the curves $\widetilde{c}(x)$ and $\widetilde{v}(x)$.
This is done in the following lemma which is analogous to
Lemma~\ref{lemma: area between c(x) and v(x)}.
\begin{lemma}{\em
\begin{equation}
  \int_0^1
  \bigl(\widetilde{v}(x)-\widetilde{c}(x)\bigr)\mathrm{d}x=\frac{C-R}{(1-R)\,a_{\text{R}}}
  \label{eq: area between c(x) and v(x) for ARA}
\end{equation}
where $\widetilde{v}$ and $\widetilde{c}$ are introduced in
\eqref{eq: tilde c and tilde v}, $C=1-p$ is the capacity of the
BEC, $R$ is the design rate of the systematic ARA ensemble, and
$a_{\text{R}}$ is defined in \eqref{eq: ARA a_L and a_R} and it
designates the average degree of the `parity-check~2' nodes when
the two edges related to the lower accumulator in
Fig.~\ref{Figure: ARA tanner graph} are not taken into account.}
\label{lemma: area between c(x) and v(x) for ARA}
\end{lemma}
\begin{proof}
The definitions of the functions $\widetilde{v}$ and
$\widetilde{c}$ in \eqref{eq: tilde c and tilde v} yield that
\begin{eqnarray}
 \int_0^1
  \bigl(\widetilde{v}(x)-\widetilde{c}(x)\bigr)\mathrm{d}x &=& \int_0^1
  \widetilde{\lambda}^{-1}(x)\mathrm{d}x - 1 +
  \int_0^1\widetilde{\rho}(1-x)\mathrm{d}x\nonumber\\
  & = & \left(1 - \int_0^1\widetilde{\lambda}(x)\mathrm{d}x\right) - 1 +
  \int_0^1\widetilde{\rho}(x)\mathrm{d}x\nonumber\\
  & = & \int_0^1\widetilde{\rho}(x)\mathrm{d}x - \int_0^1\widetilde{\lambda}(x)\mathrm{d}x
  \label{eq: first equality for integral of v(x)-c(x) for ARA}
\end{eqnarray}
where the second equality is obtained via integration by parts
(note that $\widetilde{\lambda}(0)=0$ and
$\widetilde{\lambda}(1)=1$). From \eqref{eq: L tilde}, we get
\begin{equation}
  \int_0^1 \widetilde{\lambda}(x)\mathrm{d}x = \frac{1}{\widetilde{L}'(1)} =
  \frac{p}{L'(1)} = \frac{p}{a_{\text{L}}}
  \label{eq: equation for tilde a_L}
\end{equation}
(see also \cite[Eq.~(23)]{Pfister_Sason_IT07}) where
$a_{\text{L}}$ is defined in \eqref{eq: ARA a_L and a_R} and
designates the average degree of the `punctured bit' nodes in the
Tanner graph (see Fig.~\ref{Figure: ARA tanner graph}) when the
two edges, related to the upper accumulator in Fig.~\ref{Figure:
ARA tanner graph}, are not taken into account. Similarly,
\eqref{eq: R tilde} gives
\begin{equation}
  \int_0^1 \widetilde{\rho}(x)\mathrm{d}x = \frac{1}{\widetilde{R}'(1)} =
  \frac{1-p}{R'(1)} = \frac{1-p}{a_{\text{R}}}
  \label{eq: equation for tilde a_R}
\end{equation}
(see also \cite[Eq.~(24)]{Pfister_Sason_IT07}). Substituting
\eqref{eq: equation for tilde a_L} and \eqref{eq: equation for tilde
a_R} into \eqref{eq: first equality for integral of v(x)-c(x) for
ARA} gives
\begin{eqnarray}
  \int_0^1
  \bigl(\widetilde{v}(x)-\widetilde{c}(x)\bigr)\mathrm{d}x & = &
  \frac{1-p}{a_{\text{R}}} - \frac{p}{a_{\text{L}}} \nonumber\\
  & \stackrel{(a)}{=} &
  \frac{1}{a_{\text{R}}}\,\Biggl[1-p\underbrace{\left(\frac{a_{\text{L}}
  + a_{\text{R}}}{a_{\text{L}}}\right)}_{\frac{1}{1-R}}\Biggr]\nonumber\\
  & = & \frac{1}{a_{\text{R}}}\;\frac{1-R-p}{1-R}\nonumber\\
  & = & \frac{C-R}{(1-R)\,a_{\text{R}}}
\end{eqnarray}
where $(a)$ follows since the design rate of the systematic ARA
ensemble is given by $R =
\frac{a_{\text{R}}}{a_{\text{L}}+a_{\text{R}}}$ (this equality
follows directly from Fig.~\ref{Figure: ARA tanner graph}).
\end{proof}

To continue the proof, we consider a plot similar to the one in
Figure~\ref{Figure: Fig for proof of Thm 1} with the exception
that $c(x)$ and $v(x)$ are replaced by $\widetilde{c}(x)$ and
$\widetilde{v}(x)$, respectively. Note that in this case the
horizontal line $h_0$ is reduced to the point $(1,1)$. Consider
the two sets of gray-shaded right-angled triangles. The triangles
shaded in dark gray are defined so that the height of triangle
number $i$ (counting from right to left and starting at zero) is
the vertical line $v_i$ and the slope of their hypotenuse is equal
to $\widetilde{c}'(0)=\widetilde{\rho}'(1)$. Since
$\widetilde{c}(x)$ is concave, these triangles form a subset of
the domain bounded by the curves $\widetilde{c}(x)$ and
$\widetilde{v}(x)$. The area of the $i$'th triangle in this set is
given by
\begin{equation*}
   A_i=\frac{1}{2}\;|v_i|\;\left(\frac{|v_i|}{\widetilde{\rho}'(1)}\right)
   = \frac{|v_i|^2}{2\,\widetilde{\rho}'(1)}
 \end{equation*}
where $|v_i|$ is the length of $v_i$. The second set of
right-angled triangles, which are shaded in light gray, are also
defined so that the height of the $i$'th triangle (counting from
right to left and starting at zero) is the vertical line $v_i$,
but the triangle lies to the left of $v_i$ and the slope of its
hypotenuse is equal to
\begin{equation*}
\widetilde{v}'(0) =
\left(\widetilde{\lambda}^{-1}\right)'(0)=\frac{1}{\widetilde{\lambda}'(0)}
= \frac{1}{p^2\lambda'(0)} = \frac{1}{p^2\lambda_2}
\end{equation*}
where the second equality follows since $\widetilde{\lambda}(0) =
0$ and the third equality follows from the definition of
$\widetilde{\lambda}$ in \eqref{eq: lambda tilde}. Since
$\widetilde{\lambda}$ is monotonically increasing and convex over
the interval $[0,1]$ and it satisfies $\widetilde{\lambda}(0)=0$
and $\widetilde{\lambda}(1)=1$, then it follows that $v(x) =
\widetilde{\lambda}^{-1}(x)$ is concave over this interval. Hence,
the triangles shaded in light gray also form a subset of the
domain bounded by the curves $c(x)$ and $v(x)$. The area of the
$i$'th light-gray triangle is given by
\begin{equation*}
   B_i=\frac{1}{2}\;|v_i|\;\left(|v_i|\,p^2\lambda_2\right)
   = \frac{p^2\lambda_2\,|v_i|^2}{2}
 \end{equation*}
Applying Lemma~\ref{lemma:
area between c(x) and v(x) for ARA} and the fact that the
triangles in both sets do not overlap, we get
\begin{equation}
\frac{C-R}{(1-R)\,a_{\text{R}}}\geq
\frac{1}{2}\left(\frac{1}{\widetilde{\rho}'(1)} +
p^2\lambda_2\right)\sum_{i=0}^{l-1}|v_i|^2 \label{eq: first
relationship between total area and area of triangles}
 \end{equation}
where $l$ is an arbitrary natural number. Since the sequence of
ensembles asymptotically achieves vanishing bit erasure
probability under iterative message-passing decoding, the
stability condition for systematic ARA codes (see \eqref{eq:
stability condition for ARA original version} or equivalently
\cite[Eq.~(14)]{Pfister_Sason_IT07}) implies that
\begin{equation}
  p^2\lambda_2 \leq \frac{1}{\rho'(1) + \frac{2pR'(1)}{1-p}} =
\frac{1}{\widetilde{\rho}'(1)}
  \label{eq: stability condition for ARA}
\end{equation}
where the last equality follows from \eqref{eq: rho tilde}.
Substituting \eqref{eq: stability condition for ARA} in \eqref{eq:
first relationship between total area and area of triangles} gives
\begin{equation}
  \frac{C-R}{(1-R)\,a_{\text{R}}}\geq
   p^2\lambda_2\,\sum_{i=0}^{l-1}|v_i|^2\,.
   \label{eq: second relationship between total area and area of triangles}
 \end{equation}
Let $x^{(l)}$ denote the $x$ value of the left tip of the
 horizontal line $h_l$. The value of $x^{(l)}$ satisfies the
 recursive equation
\begin{eqnarray}
  x^{(l+1)}=\widetilde{\lambda}\Bigl(1-\widetilde{\rho}\big(1-x^{(l)}\big)\Bigr),
  \; \; \forall \; l \in \naturals
\label{eq: ARA x^(l)}
\end{eqnarray}
with $ x^{(0)}=1$. As was explained above (immediately following
Lemma~\ref{lemma: x_1 as a function of previous x_1}), from
\eqref{eq: x_1 as a function of previous x_1}, \eqref{eq: ARA
initial condition for x_1}, and the monotonicity of the function
$f(x) = \widetilde{\lambda}\bigl(1-\widetilde{\rho}(1-x) \bigr)$
over the interval $[0,1]$, we get that $x^{(l)}\leq x_1^{(l)}$ for
$l\in\naturals$. The definition of $h_l$ and $v_l$ in
Figure~\ref{Figure: Fig for proof of Thm 1} implies that
\begin{equation}
  1-\widetilde{\rho}\bigl(1-x^{(l)}\bigr) =
  \widetilde{c}\bigl(x^{(l)}\bigr) = 1-\sum_{i=0}^{l}|v_i|\,.
  \label{eq: equation relating c(x^l) and sum of v_i}
\end{equation}
Starting from \eqref{eq: P_b,l for ARA} and applying the
monotonicity of $\widetilde{L}$ and $\widetilde{\rho}$ gives
\begin{eqnarray*}
    1-\sqrt{1-\frac{P_{\text{b}}^{(l-1)}}{p}} &\geq&
    \widetilde{L}\Bigl(1-\widetilde{\rho}\big(1-x_1^{(l-1)}\big)\Bigr)\nonumber\\
    &\geq&\widetilde{L}\left(1-\widetilde{\rho}\big(1-x^{(l-1)}\big)\right)\nonumber\\
    &=&\widetilde{L}\left(1-\sum_{i=0}^{l-1}|v_i|\right)
\end{eqnarray*}
where the last equality follows from \eqref{eq: equation relating
c(x^l) and sum of v_i}. Since $\widetilde{L}$ is strictly
monotonically increasing in $[0,1]$, then
\begin{equation}
  \sum_{i=0}^{l-1}|v_i| \geq
  1-\widetilde{L}^{-1}\Bigg(1-\sqrt{1-\frac{P_{\text{b}}^{(l-1)}}{p}}\Bigg)\,.
  \label{eq: equation relating P_b,l with sum of v_i}
\end{equation}
Applying the Cauchy-Schwartz inequality (as in \eqref{eq:
Cauchy-Schwarz inequality for sum of v_i}) to the RHS of
\eqref{eq: second relationship between total area and area of
triangles}, we get
\begin{eqnarray*}
   \frac{C-R}{(1-R)\,a_{\text{R}}}&\geq&
   p^2\lambda_2 \sum_{i=0}^{l-1}|v_i|^2\\
    &\geq&\frac{p^2\lambda_2}{l}\left(\sum_{i=0}^{l-1}|v_i|\right)^2\\
    &\geq&\frac{p^2\lambda_2}{l}\left(1-\widetilde{L}^{-1}
    \Bigg(1-\sqrt{1-\frac{P_{\text{b}}^{(l-1)}}{p}}\Bigg)\right)^2
\end{eqnarray*}
where the last inequality follows from \eqref{eq: equation
relating P_b,l with sum of v_i}. Since the design rate $R$ is
assumed to be a fraction $1-\varepsilon$ of the capacity of the
BEC, the above inequality gives
\begin{equation*}
  C\varepsilon \geq \frac{p^2\lambda_2\,(1-R)\,a_{\text{R}} \,
  \left(1-\widetilde{L}^{-1}\Bigg(1-\sqrt{1-
  \frac{P_{\text{b}}^{(l-1)}}{p}}\Bigg)\right)^2}{l}
\end{equation*}
where $l$ is an arbitrary natural number. Let $l$ designate the
number of iterations required to achieve an average bit erasure
probability $P_{\text{b}}$ of the systematic bits, i.e., $l$ is
the smallest integer which satisfies $P_{\text{b}}^{(l-1)}\leq
P_{\text{b}}$ (since we start counting the iterations at $l=0$).
Note that $l$ is deterministic since it refers to the smallest
number of iterations required to achieve a desired average bit
erasure probability over the ensemble. From the inequality above
and the monotonicity of $\widetilde{L}$, we obtain that
\begin{equation*}
  C\varepsilon \geq \frac{p^2\lambda_2\,(1-R)\,a_{\text{R}} \,
  \left(1-\widetilde{L}^{-1}\left(1-\sqrt{1-
  \frac{P_{\text{b}}}{p}}\right)\right)^2}{l}
\end{equation*}
which provides a lower bound on the number of iterations of the form
\begin{eqnarray}
l&\geq&\frac{p^2\lambda_2(1-R)\,a_{\text{R}} \,
  \left(1-\widetilde{L}^{-1}\left(1-\sqrt{1-
  \frac{P_{\text{b}}}{p}}\right)\right)^2}{C\varepsilon}\nonumber\\
    &=&\frac{p^2\lambda_2(1-\varepsilon)\,a_{\text{L}} \,
  \left(1-\widetilde{L}^{-1}\left(1-\sqrt{1-
  \frac{P_{\text{b}}}{p}}\right)\right)^2}{\varepsilon}
  \label{eq: first lower bound on number of iterations for ARA}
\end{eqnarray}
where the last equality follows since
$\frac{a_{\text{R}}}{a_{\text{L}}}=\frac{R}{1-R}$ (see
Fig.~\ref{Figure: ARA tanner graph}) and $R=(1-\varepsilon)C$. To
continue the proof, we derive a lower bound on
$1-\widetilde{L}^{-1}(x)$. Following the same steps which lead to
\eqref{eq: lower bound on 1-L^-1} gives the inequality
\begin{equation}
1-\widetilde{L}^{-1}(x) \geq 1-\sqrt{\frac{x}{\widetilde{L}_2}} \;
, \quad \forall \; x \geq 0
  \label{eq: lower bound on 1- tildeL^-1}
\end{equation}
where \eqref{eq: L tilde} implies that
\begin{equation}
  \widetilde{L}_2 = \frac{\widetilde{L}''(0)}{2} = \frac{p\,L''(0)}{2} =
p\,L_2\,.\label{eq: L_2 tilde}
\end{equation}
Under the assumption that $1-\sqrt{1-\frac{P_{\text{b}}}{p}} <
p\,L_2$, substituting \eqref{eq: lower bound on 1- tildeL^-1} and
\eqref{eq: L_2 tilde} in \eqref{eq: first lower bound on number of
iterations for ARA} gives
\begin{equation}
  l\geq\frac{p\lambda_2(1-\varepsilon)\, a_{\text{L}}\,
  \left(\sqrt{p\,L_2}-\sqrt{1-\sqrt{1-
  \frac{P_{\text{b}}}{p}}}
  \right)^2}{L_2\,\varepsilon}\,.
  \label{eq: second lower bound on number of iterations for ARA}
\end{equation}
Finally, the lower bound on the number of iterations in \eqref{eq:
ARA lower bound on number of iterations} follows from
\eqref{eq: second lower bound on number of iterations for ARA} by substituting $L_2 =
\frac{\lambda_2\,a_{\text{L}}}{2}$.

Considering the case where $P_{\text{b}}\rightarrow 0$ on both
sides of \eqref{eq: ARA lower bound on number of iterations} gives
\begin{equation*}
l(\varepsilon, p, P_{\text{b}}\rightarrow 0) \geq
2p^2\,(1-\varepsilon)\;\frac{L_2(\varepsilon)}{\varepsilon}\,.
\end{equation*}

\newpage
\section{Summary and Conclusions}
\label{Section: Summary and Conclusions} In this paper, we
consider the number of iterations which is required for successful
message-passing decoding of code ensembles defined on graphs. In
the considered setting, we let the block length of these ensembles
tend to infinity, and the transmission takes place over a binary
erasure channel (BEC).

In order to study the decoding complexity of these code ensembles
under iterative decoding, one needs also to take into account the
graphical complexity of the Tanner graphs of these code ensembles.
For the BEC, this graphical complexity is closely related to the
total number of operations performed by the iterative decoder. For
various families of code ensembles, Table~\ref{table: number of
iterations and graphical complexity} compares the number of
iterations and the graphical complexity which are required to
achieve a given fraction $1-\varepsilon$ (where $\varepsilon$ can
be made arbitrarily small) of the capacity of a BEC with vanishing
bit erasure probability. The results in Table~\ref{table: number
of iterations and graphical complexity} are based on lower bounds
and some achievability results which are related to the graphical
complexity of various families of code ensembles defined on graphs
(see \cite{PfisterSU_IT05, Pfister_Sason_IT07, Sason-it03,
Sason-it04}); the results related to the number of iterations are
based on the lower bounds derived in this paper.

\begin{table}[hbt]
\begin{center}
\begin{tabular}{|c|c|c|} \hline
Code & Number of decoding iterations & Graphical complexity \\
family & as function of $\varepsilon$ & as function of
$\varepsilon$
\\ \hline \hline LDPC & $\Omega\left(\frac{1}{\varepsilon}\right)$
(Theorem~\ref{theorem: number of iterations})\T \B&
$\Theta\left(\ln \frac{1}{\varepsilon}\right)$ \cite[Theorems~2.1
and 2.3]{Sason-it03}
\\ \hline
Systematic IRA & $\Omega\left(\frac{1}{\varepsilon}\right)$
(Theorem~\ref{theorem: IRA number of iterations})\T \B&
$\Theta\left(\ln
\frac{1}{\varepsilon}\right)$ \cite[Theorems~1 and 2]{Sason-it04}\\
\hline Non-systematic IRA &
$\Omega\left(\frac{1}{\varepsilon}\right)$ (Theorem~\ref{theorem:
IRA
number of iterations})\T \B & $\Theta(1)$ \cite{PfisterSU_IT05} \\
\hline Systematic ARA & $\Omega\left(\frac{1}{\varepsilon}\right)$
(Theorem~\ref{theorem: ARA number of iterations})\T \B& $\Theta(1)$ \cite{Pfister_Sason_IT07}\\
\hline
\end{tabular}
\end{center}
\caption{Number of iterations and graphical complexity required to
achieve a fraction $1-\varepsilon$ of the capacity of a BEC with
vanishing bit erasure probability under iterative message-passing
decoding.} \label{table: number of iterations and graphical
complexity}
\end{table}

Theorems~\ref{theorem: number of iterations}--\ref{theorem: IRA
number of iterations} demonstrate that for various attractive
families of code ensembles (including low-density parity-check
(LDPC) codes, systematic and non-systematic irregular
repeat-accumulate (IRA) codes, and accumulate-repeat-accumulate
(ARA) codes), the number of iterations which is required to
achieve a desired bit erasure probability scales at least like the
inverse of the gap between the channel capacity and the design
rate of the ensemble. This conclusion holds provided that the
fraction of degree-2 variable nodes in the Tanner graph does not
tend to zero as the gap to capacity vanishes (where under mild
conditions, this property is satisfied for sequences of
capacity-achieving LDPC ensembles, see
\cite[Lemma~5]{Sason-submitted07}).

When the graphical complexity of these families of ensembles is
considered, the results are less homogenous. More explicitly,
assume a sequence of LDPC codes (or ensembles) whose block length
tends to infinity, and consider the case where their transmission
takes place over a memoryless binary-input output-symmetric
channel. It follows from \cite[Theorem~2.1]{Sason-it03} that if a
fraction $1-\varepsilon$ of the capacity of this channel is
achieved with vanishing bit error (erasure) probability under ML
decoding (or any sub-optimal decoding algorithm), then the
graphical complexity of an arbitrary representation of the codes
using bipartite graphs scales at least like $\ln
\frac{1}{\varepsilon}$. For systematic IRA codes which are
transmitted over the BEC and decoded by a standard iterative
message-passing decoder, a similar result on their graphical
complexity is obtained in \cite[Theorem~1]{Sason-it04}. In
\cite[Theorem~2.3]{Sason-it03}, the lower bound on the graphical
complexity of LDPC ensembles is achieved for the BEC (up to a
small additive constant), even under iterative message-passing
decoding, by the right-regular LDPC ensembles of Shokrollahi
\cite{Shokrollahi-IMA2000}. Similarly,
\cite[Theorem~2]{Sason-it04} presents an achievability result of
this form for ensembles of systematic IRA codes transmitted over
the BEC; the graphical complexity of these ensembles scales
logarithmically with $\frac{1}{\varepsilon}$. For ensembles of
non-systematic IRA and systematic ARA codes, however, the addition
of state nodes in their standard representation by Tanner graphs
allows to achieve an improved tradeoff between the gap to capacity
and the graphical complexity; suitable constructions of such
ensembles enable to approach the capacity of the BEC with
vanishing bit erasure probability under iterative decoding while
maintaining {\em a bounded graphical complexity} (see
\cite{PfisterSU_IT05} and \cite{Pfister_Sason_IT07}). We note that
the ensembles in \cite{Pfister_Sason_IT07} have the additional
advantage of being systematic, which allows a simple decoding of
the information bits.

The lower bounds on the number of iterations in
Theorems~\ref{theorem: number of iterations}--\ref{theorem: IRA
number of iterations} become trivial when the fraction of degree-2
variable nodes vanishes. As noted in Discussion~\ref{discussion:
L_2 for capacity approaching LDPC ensembles}, for all known
capacity-approaching sequences of LDPC ensembles, this fraction
tends to $\frac{1}{2}$ as the gap to capacity vanishes. For some
ensembles of capacity approaching systematic ARA codes presented
in \cite{Pfister_Sason_IT07}, the fraction of degree-2 `punctured
bit' nodes (as introduced in Fig.~\ref{Figure: ARA tanner graph})
is defined to be zero (see \cite[Table~I]{Pfister_Sason_IT07}).
For these ensembles, the lower bound on the number of iterations
in Theorem~\ref{theorem: ARA number of iterations} is ineffective.
However, this is mainly a result of our focus on the derivation of
simple lower bounds on the number of iterations which do not
depend on the full characterization of the degree distributions of
the code ensembles. Following the proofs of Theorems~\ref{theorem:
number of iterations}~and~\ref{theorem: ARA number of iterations},
and focusing on the case where the fraction of degree-2 variable
nodes vanishes, it is possible to derive lower bounds on the
number of iterations which are not trivial even in this case;
these bounds, however, require the knowledge of the entire degree
distribution of the examined ensembles.

The simple lower bounds on the number of iterations of graph-based
ensembles, as derived in this paper, scale like the inverse of the
gap in rate to capacity and also depend on the target bit erasure
probability. The behavior of these lower bounds matches well with
the experimental results and the conjectures on the number of
iterations and complexity, as provided by Khandekar and McEliece
(see \cite{Khandekar-isit01}, \cite{Khandekar_thesis} and
\cite{McEliece_ISIT01_plenary_talk}). In
\cite[Theorem~3.5]{Khandekar_thesis}, it was stated that for LDPC
and IRA ensembles which achieve a fraction $1-\varepsilon$ of the
channel capacity of a BEC with a target bit erasure probability of
$P_{\text{b}}$ under iterative message-passing decoding, the
number of iterations grows like
$O\left(\frac{1}{\varepsilon}\right)$. In light of the outline of
the proof of this statement, as suggested in
\cite[p.~71]{Khandekar_thesis}, it implicitly assumes that the
flatness condition is satisfied for these code ensembles and also
that the target bit erasure probability vanishes; under these
assumptions, the reasoning suggested by Khandekar in
\cite[Section~3.6]{Khandekar_thesis} supports the behavior of the
lower bounds which are derived in this paper.

The matching condition for generalized extrinsic information
transfer (GEXIT) curves serves to conjecture in
\cite[Section~XI]{GEXIT} that the number of iterations scales like
the inverse of the achievable gap in rate to capacity (see also
\cite[p.~92]{Measson_Phd}); this conjecture refers to LDPC
ensembles whose transmission takes place over a general memoryless
binary-input output-symmetric (MBIOS) channel. Focusing on the
BEC, the derivation of the lower bounds on the number of
iterations (see Section~\ref{Section: Proofs of Main Results})
makes the heuristic reasoning of this scaling rigorous. It also
extends the bounds to various graph-based code ensembles (e.g.,
IRA and ARA ensembles) under iterative message-passing decoding,
and makes them universal for the BEC in the sense that they are
expressed in terms of some basic parameters of the ensembles which
include the fraction of degree-2 variable nodes, the target bit
erasure probability and the asymptotic gap between the channel
capacity and the design rate of the ensemble (but the bounds here
do not depend explicitly on the degree distributions of the code
ensembles). An interesting and challenging direction which calls
for further research is to extend these lower bounds on the number
of iterations for general MBIOS channels; as suggested in
\cite[Section~~XI]{GEXIT}, a consequence of the matching condition
for GEXIT curves has the potential to lead to such lower bounds on
the number of iterations which also scale like the inverse of the
gap to capacity for general MBIOS channels.

\subsection*{Acknowledgment} This research work was supported by the Israel Science
Foundation (grant no. 1070/07). The work was initiated during a
visit of the first author at EPFL in Lausanne, Switzerland, and it
benefited from a short unpublished write-up which was jointly
written by S. Dusad, C. Measson, A. Montanari and R. Urbanke. This
included preliminary steps towards the derivation of a lower bound
on the number of iterations for LDPC ensembles. The first author
also wishes to acknowledge H. D. Pfister for various discussions
prior to this work on accumulate-repeat-accumulate codes; the
`graph reduction' principle presented in \cite{Pfister_Sason_IT07}
for the binary erasure channel was helpful in the derivation of
Theorems~\ref{theorem: ARA number of iterations} and~\ref{theorem:
IRA number of iterations}.

\section*{\center{Appendices}}
\setcounter{section}{0}
\renewcommand{\thesection}{Appendix~\Roman{section}}
\renewcommand{\theequation}{\Roman{section}.\arabic{equation}}
\renewcommand{\thelemma}{\Roman{section}.\arabic{lemma}}
\renewcommand{\thesubsection}{\Roman{section}.\Alph{subsection}}
\setcounter{equation}{0} \setcounter{lemma}{0}
\setcounter{subsection}{0} 
\section{Proof of Proposition~\ref{proposition: ARA number of iterations turbo}}
\begin{figure}[hbt]
\begin{center}
\epsfig{file=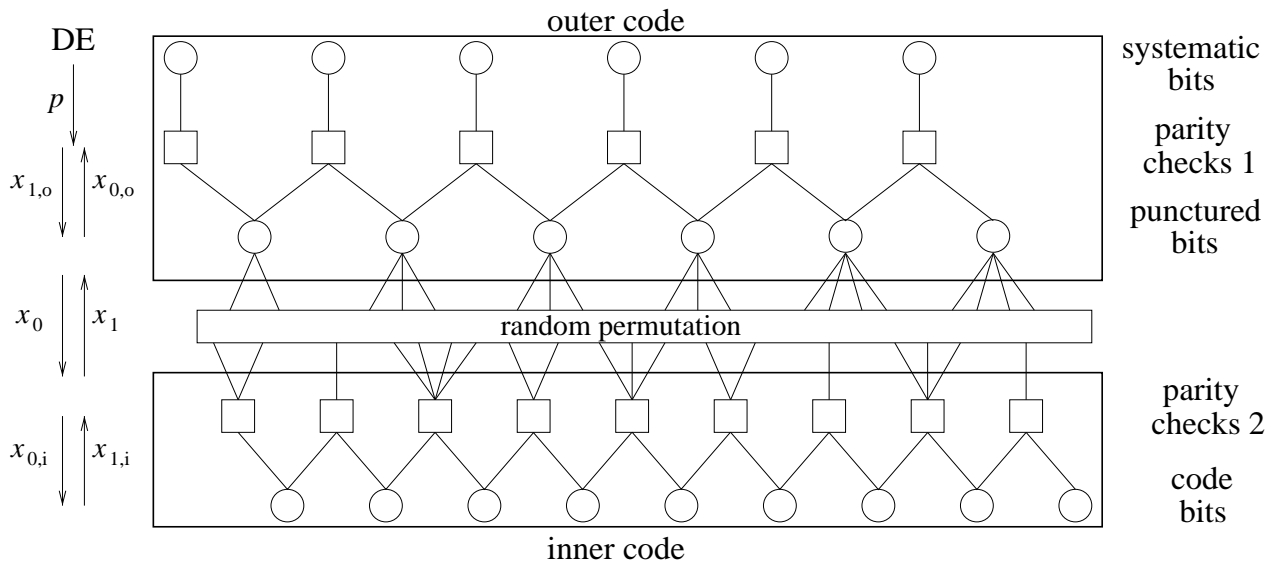,scale=0.85}
\end{center}
\caption{\label{Figure: ARA turbo-like tanner graph} Tanner graph
of a systematic accumulate-repeat-accumulate (ARA) code for
turbo-like decoding as an interleaved and serially concatenated
code.}
\end{figure}

\label{Appendix: Proof of proposition ARA number of iterations
turbo} We begin the proof by considering an iterative decoder of
systematic ARA codes by viewing them as interleaved and serially
concatenated codes. The outer code of the systematic ARA code
consists of the first accumulator which operates on the systematic
bits (see the upper zigzag in Fig.~\ref{Figure: ARA tanner graph}),
followed by the irregular repetition code. The inner code consists
of the irregular SPC code, followed by the second accumulator (see
the lower zigzag in Fig.~\ref{Figure: ARA tanner graph}). These two
constituent codes are joined by an interleaver which permutes the
repeated bits at the output of the outer code before they are used
as input to the inner encoder; for the considered ARA ensemble, we
assume that the interleaver is chosen uniformly at random over all
interleavers of the appropriate length. The turbo-like decoding
algorithm is based on iterating extrinsic information between
bitwise MAP decoders of the two constituent codes (see e.g.,
\cite{Serially_Concatenated_codes}). Each decoding iteration begins
with an extrinsic bitwise MAP decoding for each non-systematic
output bit of the outer code (these are the bits which serve as
input to the inner code) based on the information regarding these
bits received from the extrinsic bitwise MAP decoder of the inner
code in the previous iteration and the information on the systematic
bits received from the communication channel. In the second stage of
the iteration, this information is passed from the outer decoder to
an extrinsic bitwise MAP decoder of the inner code and is used as
a-priori knowledge for decoding the input bits of the inner code. A
Tanner graph for turbo-like decoding of systematic ARA codes is
presented in Figure~\ref{Figure: ARA turbo-like tanner graph}.
Considering the asymptotic case where the block length tends to
infinity, we denote the probability of erasure messages from the
outer decoder to the inner decoder and vice versa at the $l$'th
decoding iteration by $x_0^{(l)}$ and $x_1^{(l)}$, respectively.
Keeping in line with the notation in the proofs of
Theorems~\ref{theorem: number of iterations}~and~\ref{theorem: ARA
number of iterations}, we begin counting the iterations at $l=0$.
Since there is no a-priori information regarding the non-systematic
output bits of the outer decoder (which are permuted to form the
input bits of the inner decoder, as shown in Fig.~\ref{Figure: ARA
turbo-like tanner graph}) we have
\begin{equation}
  x_0^{(-1)} = x_1^{(-1)} = 1.
\label{eq: intial condition for turbo-like message erasure
probability}
\end{equation}

We now turn to calculate the erasure probability $x_0^{(l)}$ in an
extrinsic bitwise MAP decoding of non-systematic output bits of
the outer code, given that the a-priori erasure probability of
these bits is $x_1^{(l-1)}$. To this end, we consider the Tanner
graph of the outer code, shown in the top box of
Figure~\ref{Figure: ARA turbo-like tanner graph}. We note that
this Tanner graph contains no cycles, and therefore bitwise MAP
decoding of this code can be performed by using the standard
iterative message-passing decoding algorithm until a fixed-point
is reached. In such a decoder which operates on the Tanner graph
of the outer code, messages are transferred between the `punctured
bit' and the `parity-check~1' nodes of the graph. Let us denote by
$x_{0,\text{o}}(x)$ the probability of erasure in messages from
the `punctured bit' nodes to the `parity-check 1' nodes at the
fixed point of the iterative decoding algorithm, when the a-priori
erasure probability of the output bits is $x$. Similarly, we
denote by $x_{1,\text{o}}(x)$ the erasure probability in messages
from the 'parity-check 1' nodes to the 'punctured bit' nodes at
the fixed point, where $x$ is the a-priori erasure probability of
the non-systematic output bits. Based on the structure of the
Tanner graph, we have
\begin{equation}
  x_{0,\text{o}}(x) = x_{1,\text{o}}(x)\cdot L(x)
  \label{eq: fixed point of x_0,o(x) as function of x_1,o(x)}
\end{equation}
and
\begin{equation}
  x_{1,\text{o}}(x) = 1-(1-p)\big(1-x_{0,\text{o}}(x)\big)
  \label{eq: fixed point of x_1,o(x) as function of x_0,o(x)}
\end{equation}
where $L$ is defined in \eqref{eq: ARA L} and it forms the degree
distribution of the `punctured bit' nodes from the node
perspective, and $p$ denotes the erasure probability of the BEC.
Substituting \eqref{eq: fixed point of x_0,o(x) as function of
x_1,o(x)} into \eqref{eq: fixed point of x_1,o(x) as function of
x_0,o(x)} gives
\begin{equation}
x_{1,\text{o}}(x) = \frac{p}{1-(1-p)L(x)}\,.
  \label{eq: fixed point of x_1,o(x)}
\end{equation}
Therefore, the structure of the Tanner graph of the outer code implies that the
erasure probability $x_0^{(l)}$ in messages from the outer decoder
to the inner decoder at iteration number $l$ of the turbo-like decoding algorithm is given by
\begin{eqnarray}
x_0^{(l)} &=&
\Bigl(x_{1,\text{o}}\big(x_1^{(l-1)}\big)\Bigr)^2 \lambda\big(x_1^{(l-1)}\big)\nonumber\\
 &=&\left(\frac{p}{1-(1-p)L\big(x_1^{(l-1)}\big)}\right)^2
    \lambda\big(x_1^{(l-1)}\big)\nonumber\\
&=&\widetilde{\lambda}\big(x_1^{(l-1)}\big)
  \label{eq: x_0^(l) for turbo decoding}
\end{eqnarray}
where the second equality relies on \eqref{eq: fixed point of
x_1,o(x)}, and $\widetilde{\lambda}$ is the tilted degree
distribution which results from graph reduction (see \eqref{eq:
lambda tilde}). We now employ a similar technique to calculate the
erasure probability $x_1^{(l)}$ in an extrinsic bitwise MAP
decoding of input bits of the inner code, given that the a-priori
erasure probability of these bits is $x_0^{(l)}$. Since the Tanner
of the inner code is also cycle-free (see the lower box in
Figure~\ref{Figure: ARA turbo-like tanner graph}), extrinsic
bitwise MAP decoding can be done by using the iterative decoder
operating on the Tanner graph of the inner code. We denote by
$x_{0,\text{i}}(x)$ the erasure probability of messages from the
`parity check 2' nodes to the `code bit' nodes at the fixed point
of the iterative decoding algorithm when $x$ is the a-priori
erasure probability of the input bits. Similarly,
$x_{1,\text{i}}(x)$ designates the erasure probability of messages
from the `code bit' nodes to the `parity check 2' nodes at the
fixed point of the decoding algorithm, when $x$ is the a-priori
erasure probability of the input bits. The structure of the Tanner
graph implies that
\begin{equation}
  x_{0,\text{i}}(x) = 1-\big(1-x_{1,\text{i}}(x)\big)R(1-x)
  \label{eq: fixed point of x_0,i(x) as function of x_1,i(x)}
\end{equation}
and
\begin{equation}
  x_{1,\text{i}}(x) = p\,x_{0,\text{i}}(x)
  \label{eq: fixed point of x_1,i(x) as function of x_0,i(x)}
\end{equation}
where $R$ is defined in \eqref{eq: ARA R}. Substituting \eqref{eq:
fixed point of x_0,i(x) as function of x_1,i(x)} into \eqref{eq:
fixed point of x_1,i(x) as function of x_0,i(x)} gives
\begin{equation}
  x_{1,\text{i}}(x) = \frac{p\bigl(1-R(1-x)\bigr)}{1-p\,R(1-x)}\,.
  \label{eq: fixed point of x_1,i(x)}
\end{equation}
Therefore, the erasure probability $x_1^{(l)}$ in messages from
the inner decoder to the outer decoder at iteration number $l$ of
the turbo-like decoding algorithm is given by
\begin{eqnarray}
&& x_1^{(l)} = 1-\Big(1-x_{1,\text{i}}\big(x_0^{(l)}\big)\Big)^2
\rho\big(1-x_0^{(l)}\big) \nonumber\\
&& \hspace*{0.7cm} =
1-\left(1-\frac{p\Bigl(1-R\big(1-x_0^{(l)}\big)\Bigr)}{1-p\,R\big(1-x_0^{(l)}\big)}\right)^2
\rho\big(1-x_0^{(l)}\big)\nonumber\\
&& \hspace*{0.7cm} =
1-\left(\frac{1-p}{1-p\,R\big(1-x_0^{(l)}\big)}\right)^2
\rho\big(1-x_0^{(l)}\big)\nonumber\\
&& \hspace*{0.7cm} = 1-\widetilde{\rho}\big(1-x_0^{(l)}\big)
  \label{eq: x_1^(l) for turbo decoding}
\end{eqnarray}
where the second equality relies on \eqref{eq: fixed point of
x_1,i(x)}, and $\widetilde{\rho}$ is the tilted degree
distribution resulting from graph reduction (see \eqref{eq: rho
tilde}). Combining \eqref{eq: intial condition for turbo-like
message erasure probability}, \eqref{eq: x_0^(l) for turbo
decoding} and \eqref{eq: x_1^(l) for turbo decoding} gives
\begin{eqnarray}
&& x_0^{(0)} =
\widetilde{\lambda}\big(x_1^{(-1)}\big)=\widetilde{\lambda}(1)=1\,,\nonumber\\
&& x_0^{(l)} =
\widetilde{\lambda}\Big(1-\widetilde{\rho}\big(1-x_0^{(l-1)}\big)\Big)\,
,\quad\quad l\in\naturals\,.
  \label{eq: x_0^(l) for turbo decoding with initial condition}
\end{eqnarray}
Observing the proof of Theorem~\ref{theorem: ARA number of
iterations}, we note that $x_0^{(l)} = x^{(l)}$ for all
$l=0,1,\ldots$, where is the $x^{(l)}$ value at the left tip of the
horizontal line $h_l$ in Figure~\ref{Figure: Fig for proof of Thm 1}
(see Eq.~\eqref{eq: ARA x^(l)} on page~\pageref{eq: ARA x^(l)}).

Let $P_{\text{b}}^{(l)}$ designate the average erasure probability
of the systematic bits at the end of the $l$'th iteration of the
turbo-like decoder. From the definition of the turbo-like decoding
algorithm, $P_{\text{b}}^{(l)}$ is the erasure probability of
bitwise MAP decoding for the input bits to the outer code, given
that the a-priori erasure probability of the output bits of this
code is given by $x_1^{(l)}$. Based of the structure of the Tanner
graph of the outer code in Figure~\ref{Figure: ARA turbo-like tanner
graph}, we get
\begin{equation}
P_{\text{b}}^{(l)} =
p\,\left[1-\Big(1-x_{0,\text{o}}\big(x_1^{(l)}\big)\Big)^2\right]
  \label{eq: initial expression for P_b,l for turbo-like decoding of ARA}
\end{equation}
where $x_{0,\text{o}}(x)$ in the fixed point erasure probability
of messages from the `punctured bit' nodes to the `parity-check 1'
nodes in the case that the a-priori erasure probability of the
non-systematic output bits of the code is $x$. Substituting
\eqref{eq: fixed point of x_1,o(x) as function of x_0,o(x)} in
\eqref{eq: fixed point of x_0,o(x) as function of x_1,o(x)} gives
\begin{equation*}
  x_{0,\text{o}}(x) = \frac{p\,L(x)}{1-(1-p)L(x)}\,.
\end{equation*}
Substituting the above equality into \eqref{eq: initial expression
for P_b,l for turbo-like decoding of ARA}, we have
\begin{eqnarray*}
P_{\text{b}}^{(l)} &=&
p\,\left[1-\bigg(1-\frac{p\,L\big(x_1^{(l)}\big)}{1-(1-p)L\big(x_1^{(l)}\big)}\bigg)^2\right]\nonumber\\
&=&
p\,\left[1-\Big(1-\widetilde{L}\big(x_1^{(l)}\big)\Big)^2\right]\nonumber\\
&=&
p\,\left[1-\bigg(1-\widetilde{L}\Big(1-\widetilde{\rho}\big(1-x_0^{(l)}\big)\Big)\bigg)^2\right]
\end{eqnarray*}
where the second equality follows from the definition of
$\widetilde{L}$ in \eqref{eq: L tilde} and the third equality
relies on \eqref{eq: x_1^(l) for turbo decoding}. Using simple
algebra, the above expression gives
\begin{equation}
1-\sqrt{1-\frac{P_{\text{b}}^{(l)}}{p}} =
\widetilde{L}\Big(1-\widetilde{\rho}\big(1-x_0^{(l)}\big)\Big)\,.
  \label{eq: expression for P_b,l for turbo-like decoding of ARA}
\end{equation}
Hence, the lower bound on the average erasure probability of the
systematic bits at the end of the $l$'th iteration of the standard
iterative decoder for ARA codes in Lemma~\ref{lemma: P_b,l for ARA}
is satisfied (with equality) also for the turbo-like decoder.

Let $l$ designate the required number of iterations for the
turbo-like decoder to achieve an average erasure probability
$P_{\text{b}}$ of the systematic bits. Since we start counting the
iterations at zero, \eqref{eq: expression for P_b,l for turbo-like
decoding of ARA} implies that $l$ is the smallest natural number
which satisfies
\begin{equation*}
1-\sqrt{1-\frac{P_{\text{b}}}{p}} \geq
\widetilde{L}\Big(1-\widetilde{\rho}\big(1-x_0^{(l-1)}\big)\Big)\,.
\end{equation*}
However, this is exactly the quantity for which we calculated the
lower bound in the proof of Theorem~\ref{theorem: ARA number of
iterations} (see Lemmas~\ref{lemma: x_1 as a function of previous
x_1}~and~\ref{lemma: P_b,l for ARA} and Eq.~\eqref{eq: ARA initial
condition for x_1}). Therefore, we conclude that the lower bound
on the number of iterations $(l)$ in Theorem~\ref{theorem: ARA
number of iterations} holds also when the considered turbo-like
decoding algorithm is employed to decode the systematic ARA codes
as interleaved and serially concatenated codes.

\vspace*{-0.5cm}
\section{}
\setcounter{equation}{0} \setcounter{lemma}{0}
\setcounter{subsection}{0}
\subsection{Proof of Lemma~\ref{lemma: x_1 as a function of previous x_1}}
\label{Appendix: Proof of lemma: x_1 as a function of previous
x_1} The proof of Lemma~\ref{lemma: x_1 as a function of previous
x_1} is based on the DE equations in \eqref{eq: ARA DE equations}
for systematic ARA ensembles. From the DE equations for
$x_2^{(l)}$ and $x_3^{(l)}$, we have
\begin{eqnarray*}
  x_3^{(l)} &=& p\,x_2^{(l)} \\
  &=&
  p\left[1-R\left(1-x_1^{(l)}\right)\left(1-x_3^{(l-1)}\right)\right]\\
  &\geq&
  p\left[1-R\left(1-x_1^{(l)}\right)\left(1-x_3^{(l)}\right)\right]
\end{eqnarray*}
where the inequality follows since the decoding process does not
add erasures, so $x_i^{(l)}$ is monotonically decreasing with $l$
(for $i=0,1,\ldots,5$). This gives
\begin{equation*}
1-x_3^{(l)}\leq
1-p\left[1-R\left(1-x_1^{(l)}\right)\left(1-x_3^{(l)}\right)\right]
\end{equation*}
and
\begin{equation}
1-x_3^{(l)}\leq \frac{1-p}{1-pR\left(1-x_1^{(l)}\right)}.
\label{eq: inequality for 1-x_3}
\end{equation}
Substituting \eqref{eq: inequality for 1-x_3} into the DE equation
for $x_4^{(l)}$ (see \eqref{eq: ARA DE equations}) gives
\begin{eqnarray}
x_4^{(l)} &=&
1-\left(1-x_3^{(l)}\right)^2\rho\left(1-x_1^{(l)}\right)\nonumber\\
&\geq&1-\left(\frac{1-p}{1-pR\left(1-x_1^{(l)}\right)}\right)^2\rho\left(1-x_1^{(l)}\right)\nonumber\\
&=& 1-\widetilde{\rho}\left(1-x_1^{(l)}\right) \label{eq:
inequality of x_4}
\end{eqnarray}
where $\widetilde{\rho}$ is defined in \eqref{eq: rho tilde}. From
\eqref{eq: ARA DE equations}, we get
\begin{eqnarray*}
  x_5^{(l)} &=& x_0^{(l)}L\left(x_4^{(l)}\right)\\
  &=&\left[1-\left(1-x_5^{(l-1)}\right)(1-p)\right]L\left(x_4^{(l)}\right)\\
  & \geq & \left[1-\left(1-x_5^{(l)}\right)(1-p)\right]L\left(x_4^{(l)}\right)
\end{eqnarray*}
where the inequality follows from the monotonicity of
$\{x_5^{(l)}\}$. Solving for $1-x_5^{(l)}$ gives
\begin{equation}
  1-x_5^{(l)} \leq
  \frac{1-L\left(x_4^{(l)}\right)}{1-(1-p)L\left(x_4^{(l)}\right)}\,.
  \label{eq: inequality for 1-x_5}
\end{equation}
Substituting \eqref{eq: inequality for 1-x_5} into the DE equation
for $x_0^{(l)}$ in \eqref{eq: ARA DE equations}, we have
\begin{eqnarray*}
x_{0}^{(l)} & = & 1-\left(1-x_5^{(l-1)}\right)(1-p)\\
& \geq &
1-\frac{(1-p)\left[1-L\left(x_4^{(l-1)}\right)\right]}{1-(1-p)L\left(x_4^{(l-1)}\right)}\\
& = & \frac{p}{1-(1-p)L\left(x_4^{(l-1)}\right)}\,.
\end{eqnarray*}
Substituting the inequality above into the DE equation for
$x_1^{(l)}$ gives
\begin{eqnarray}
x_1^{(l)} &=&
\left(x_0^{(l)}\right)^2\lambda\left(x_4^{(l-1)}\right)\nonumber\\
&\geq&
\left(\frac{p}{1-(1-p)L\left(x_4^{(l-1)}\right)}\right)^2\lambda\left(x_4^{(l-1)}\right)\nonumber\\
&=&\widetilde{\lambda}\left(x_4^{(l-1)}\right)\label{eq:
inequality of x_1}
\end{eqnarray}
where $\widetilde{\lambda}$ is defined in \eqref{eq: lambda
tilde}. Finally \eqref{eq: x_1 as a function of previous x_1}
follows from \eqref{eq: inequality of x_4} and \eqref{eq:
inequality of x_1} and the monotonicity of $\widetilde{\lambda}$
over the interval $[0,1]$.

\subsection{Proof of Lemma~\ref{lemma: P_b,l for ARA}}
\label{Appendix: Proof of lemma: P_b,l for ARA}
  From the structure of the Tanner graph of systematic ARA codes
  (see Fig.~\ref{Figure: ARA tanner graph}) and the DE equation for
  $x_5^{(l)}$ in \eqref{eq: ARA DE equations} we get
  \begin{eqnarray}
    P_{\text{b}}^{(l)} &=&
    p\left[1-\left(1-x_5^{(l)}\right)^2\right]\nonumber\\
    &=&p\left[1-\left(1-x_0^{(l)}L\left(x_4^{(l)}\right)\right)^2\right]\,.
    \label{eq: First equality for P_b,l}
  \end{eqnarray}
The DE equation \eqref{eq: ARA DE equations} for $x_1^{(l)}$ and
\eqref{eq: lambda tilde} imply that
\begin{eqnarray*}
  \left(x_0^{(l)}\right)^2 &=&
  \frac{x_1^{(l)}}{\lambda\left(x_4^{(l-1)}\right)}\\
  &=& \frac{x_1^{(l)}\,p^2}{\widetilde{\lambda}\left(x_4^{(l-1)}\right) \,
  \left[1-(1-p)\,L\left(x_4^{(l-1)}\right)\right]^2}\\
  &\geq& \left(\frac{p}{1-(1-p)\,L\left(x_4^{(l-1)}\right)}\right)^2
\end{eqnarray*}
where the last inequality follows from \eqref{eq: inequality of
x_1}. Taking the square root on both sides of the above inequality
gives
\begin{equation}
  x_0^{(l)} \geq \frac{p}{1-(1-p)\,L\left(x_4^{(l-1)}\right)}\,.
  \label{eq: first inequality for x_0^l}
\end{equation}
Substituting \eqref{eq: first inequality for x_0^l} in \eqref{eq:
First equality for P_b,l}, we get
\begin{eqnarray}
P_{\text{b}}^{(l)} &\geq&
p\left[1-\left(1-\frac{p\,L\left(x_4^{(l)}\right)}
{1-(1-p)\,L\left(x_4^{(l-1)}\right)}\right)^2\right]\nonumber\\
&\geq& p\left[1-\left(1-\frac{p\,L\left(x_4^{(l)}\right)}
{1-(1-p)\,L\left(x_4^{(l)}\right)}\right)^2\right] \label{eq: second
inequality for P_b,l}
\end{eqnarray}
where the second inequality above follows since the decoding
process does not add erasures so $x_4^{(l)}\leq x_4^{(l-1)}$, and
from the monotonicity of $L$ over $[0,1]$. Applying the definition
of $\widetilde{L}$ in \eqref{eq: L tilde} to the RHS of \eqref{eq:
second inequality for P_b,l} gives
\begin{eqnarray}
P_{\text{b}}^{(l)}& \geq &p\left[1-\left(1-\widetilde{L}\left(x_4^{(l)}\right)\right)^2\right]\nonumber\\
&\geq&
p\left\{1-\left[1-\widetilde{L}\left(1-\rho\left(x_1^{(l)}\right)\right)\right]^2\right\}
\label{eq: Final inequality for P_b,l}
\end{eqnarray}
where the last inequality follows from \eqref{eq: inequality of
x_4}. Finally, \eqref{eq: P_b,l for ARA} follows directly from
\eqref{eq: Final inequality for P_b,l}.

\end{document}